\documentclass[12pt]{article}

\usepackage{amsmath}
\usepackage{amsfonts}
\usepackage{graphicx}
\usepackage[a4paper, total={18cm, 23cm}]{geometry}
\usepackage{multirow}
\usepackage{tikz}
\usetikzlibrary{arrows}
\usetikzlibrary{decorations.markings}
\usepackage{color}
\usepackage{soul}
\usepackage{colortbl}
\usepackage{rotating}
\usepackage{natbib}

\definecolor{darkgreen}{rgb}{0.05, 0.5, 0.05}
\newcommand{\indep}{\perp \!\!\! \perp}
\newcommand{\notindep}{\not \! \perp \!\!\! \perp}

\begin{document}

\title{\textbf{\Large{Using Instruments for Selection to Adjust for Selection Bias in Mendelian Randomization}}}
\author{Apostolos Gkatzionis$^{1,2}$, Eric J. Tchetgen Tchetgen$^{3}$, Jon Heron$^{1,2}$, \\
Kate Northstone$^{2}$, Kate Tilling$^{1,2}$}
\date{\begin{small}$^1$MRC Integrative Epidemiology Unit, University of Bristol, UK.
$^2$Population Health Science Institute, Bristol Medical School, University of Bristol, UK.
$^3$Department of Statistics and Data Science, Wharton School, University of Pennsylvania, USA.
\end{small}}
\maketitle

\abstract{Selection bias is a common concern in epidemiologic studies. In the literature, selection bias is often viewed as a missing data problem. Popular approaches to adjust for bias due to missing data, such as inverse probability weighting, rely on the assumption that data are missing at random and can yield biased results if this assumption is violated. In observational studies with outcome data missing not at random, Heckman's sample selection model can be used to adjust for bias due to missing data. In this paper, we review Heckman's method and a similar approach proposed by Tchetgen Tchetgen and Wirth (2017). We then discuss how to apply these methods to Mendelian randomization analyses using individual-level data, with missing data for either the exposure or outcome or both. We explore whether genetic variants associated with participation can be used as instruments for selection. We then describe how to obtain missingness-adjusted Wald ratio, two-stage least squares and inverse variance weighted estimates. The two methods are evaluated and compared in simulations, with results suggesting that they can both mitigate selection bias but may yield parameter estimates with large standard errors in some settings. In an illustrative real-data application, we investigate the effects of body mass index on smoking using data from the Avon Longitudinal Study of Parents and Children.}


\section{Introduction}

Mendelian randomization (MR) uses genetic data to assess the causal relationship between a modifiable exposure and an outcome of interest \citep{DaveySmith2003, Burgess2015book}. MR is an application of instrumental variables analysis where genetic variants are used as instruments. The instrumental variable framework allows MR to account for unobserved confounding, which is a primary concern in other types of observational studies. However, just like any other epidemiological study, MR analyses remain susceptible to selection bias. There are several examples where this may occur. Studies of disease progression can be affected by selection bias, due to only observing disease progression traits on individuals who have developed the disease \citep{Mitchell2022}. Survivor bias can have an impact on MR analyses conducted on elderly cohorts. And large-scale genetic datasets that are commonly used for MR, such as the UK Biobank, can yield biased results if the selection of participants into the dataset is not representative of the population from which they are selected \citep{Fry2017, Batty2020, Pirastu2021}.

In this paper, we view selection bias as a missing data problem. Common approaches to adjust for missing data include inverse probability weighting (IPW) and multiple imputation; several authors have already considered the use of IPW to adjust for selection bias in MR \citep{Canan2017, Hughes2017, Munafo2018, Gkatzionis2018}. These methods attempt to model the pattern of missingness or the distribution of missing values using fully observed variables, and thus rely on the assumption that data are missing at random (MAR). This assumption is questionable in some applications. For example, in the early days of the Covid-19 pandemic, only individuals in high-risk groups or those who exhibited Covid-19 symptoms were tested for the virus. Asymptomatic patients from low risk groups were not tested and could not be classified as having Covid-19. Therefore, in many Covid-19 studies conducted in 2020, participation depended on developing Covid-19-related symptoms \citep{Griffith2020}, and developing symptoms was in turn associated with Covid-19 outcomes (e.g. disease severity) resulting in data missing not at random (MNAR).

In this paper, we discuss how to adjust for selection bias in instrumental variable and Mendelian Randomization studies with individual-level MNAR data. Our approach is based on Heckman's sample selection model \citep{Heckman1979}, which was developed for observational studies in econometrics, as well as a recently proposed alternative method by \cite{Tchetgen2017}. The main idea is to identify a variable that is observed for all study participants and associated with selection into a study but not associated with the exposure and outcome of interest in the underlying population. Such a variable has been called an instrumental variable for selection, and can be used as an external source of information about the selection process, with which to adjust for selection bias. Despite being commonly used in econometrics, this approach has been underappreciated in genetic epidemiology.

We first present the method in the framework regression analyses in Section 2, starting from Heckman's original model for linear regression and then considering the more general setting of Tchetgen Tchetgen and Wirth. We then discuss our adaptation for MR studies in Section 3. Section 4 contains a range of simulations that were conducted to evaluate the two methods' performance, and in Section 5 we implement a real-data application, assessing the effect of Body Mass Index (BMI) on smoking status and intensity, using data from the Avon Longitudinal Study of Parents and Children (ALSPAC).

\section{Instruments for Selection in Regression}

\subsection{Instrumental Variable Assumptions for Selection}

In this section, our interest lies in estimating the association between a vector of covariates $X$ and an outcome $Y$, based on a sample $(X_i, Y_i), i = 1, \dots, n$. In particular, we consider the case where covariate values $X_i$ are observed for all individuals in the sample, but outcome values $Y_i$ are not. Let $R$ be a selection indicator, such that if $R_i = 1$ we observe $Y_i$ and if $R_i = 0$ we do not. The observed data are $(X_i, R_i, R_i Y_i)$.

One approach to account for the missing outcome values in this setting is to use inverse probability weighting. IPW specifies a regression model for the selection indicator $R$ and then weights each individual by the inverse of the probability of observing their outcome, in order to account for individuals with similar characteristics, for which the outcome is not observed. Standard IPW relies on a MAR assumption, which takes the form $\mathbb{P} (R | X, Y) = \mathbb{P} (R | X)$ for the simple setting described above. If additional variables that affect missingness are available, these can be included in the model for $R$. Under the MAR assumption, if the weighting model is correctly specified, weighting by IPW is able to eliminate selection bias. 

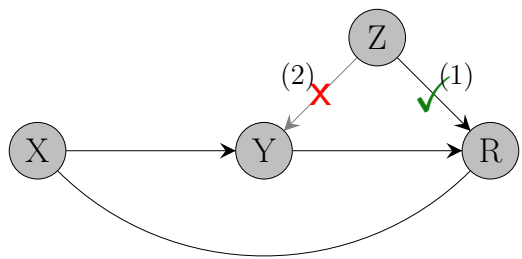
\begin{figure}[h]
\begin{center}
\begin{tikzpicture}[scale=0.8, every node/.style={scale=0.8}]
\draw [decoration={markings,mark=at position 1 with
    {\arrow[scale=2,>=stealth]{>}}},postaction={decorate}] (-3.5, 0) -- (-0.5, 0);
\draw [gray, decoration={markings,mark=at position 1 with
    {\arrow[scale=2,>=stealth]{>}}},postaction={decorate}] (1.65, 1.65) -- (0.35, 0.35);
\draw [decoration={markings,mark=at position 1 with
    {\arrow[scale=2,>=stealth]{>}}},postaction={decorate}] (2.35, 1.65) -- (3.65, 0.35);
\draw [decoration={markings,mark=at position 1 with
    {\arrow[scale=2,>=stealth]{>}}},postaction={decorate}] (0.5, 0) -- (3.5, 0);
\draw [decoration={markings,mark=at position 1 with
    {\arrow[scale=2,>=stealth]{>}}},postaction={decorate}] (-3.65, -0.35) to [out = 315, in = 225] (3.65, -0.35);
\draw[fill = lightgray] (0, 0) circle (0.5cm);
\draw[fill = lightgray] (-4, 0) circle (0.5cm);
\draw[fill = lightgray] (4, 0) circle (0.5cm);
\draw[fill = lightgray] (2, 2) circle (0.5cm);
\node (a) at (0, 0) {\Large{Y}};
\node (b) at (-4, 0) {\Large{X}};
\node (c) at (4, 0) {\Large{R}};
\node (d) at (2, 2) {\Large{Z}};
\node (f) at (3, 1) {\textcolor{darkgreen}{\huge{\checkmark}}};
\node (h) at (1, 1) {\textcolor{red}{\huge{\textsf{x}}}};
\node (i) at (0.6, 1.3) {\large{(2)}};
\node (k) at (3.4, 1.3) {\large{(1)}};
\end{tikzpicture}
\end{center}
\caption{An illustration of the assumptions characterizing an instrumental variable for selection in observational studies.} 
\label{obs.dag}
\end{figure}

Now suppose that data are missing not at random (MNAR), whence $\mathbb{P} (R | X, Y)$ depends on $Y$. Following the relevant literature, we let $Z$ denote an additional variable that is observed for all individuals in the sample and is known to be associated with the selection indicator $R$. The variable $Z$ is called an instrumental variable for selection if it satisfies two basic conditions:
\begin{enumerate}
	\item $Z \notindep R | X$: the instrumental variable is associated with outcome missingness conditional on the observed covariates. 
	\item $Z \indep Y | X$: the instrumental variable is independent of the outcome in the underlying population, conditional on observed covariates.
\end{enumerate}
These two assumptions are visualized in Figure~\ref{obs.dag}. The first assumption resembles the relevance assumption of traditional instrumental variable analyses. The second assumption is similar to an exclusion restriction assumption; note that it does not prevent $Z$ from exerting an effect on the covariates $X$ or vice versa, nor does it prevent $Z$ from affecting the outcome $Y$ through its effect on $X$. Formally, the first assumption is testable but the second assumption is not. Together, they establish that $Z$ can be used as an external source of information about the selection process, with which to adjust for selection bias.

\subsection{Heckman's Selection Model}

The idea of using an instrumental variable for selection in order to adjust for selection bias was first proposed by Heckman \citep{Heckman1976, Heckman1979}. The modelling approach developed in these papers has become known as Heckman's sample selection model and is widely used in econometrics. Heckman considered the case of a normally distributed outcome,
\begin{equation}
	Y_i | X_i = X_i^T \beta + \epsilon_i \;\;\;\;\; , \;\;\;\;\; \epsilon_i \sim N(0, \sigma_1^2) \label{linear}
\end{equation}
that is observed or missing according to a latent continuous process,
\begin{equation}
	\tilde{Y}_i | \tilde{X}_i = \tilde{X}_i^T \gamma + \tilde{\epsilon}_i \;\;\;\;\; , \;\;\;\;\; \tilde{\epsilon}_i \sim N(0, \sigma_2^2) \label{heck.latent}
\end{equation}
where $\tilde{X}$ represents variables that affect $\tilde{Y}$, and the error terms are correlated: $\text{Cov} (\epsilon_i, \tilde{\epsilon}_i) = \sigma_{12}$. The vector $\tilde{X}$ must include $Z$ and may also include some of the covariates $X$ or other fully observed variables. The inclusion of $Z$ in \eqref{heck.latent} but not in \eqref{linear} is important to avoid issues of collinearity when fitting the model \citep{Puhani2000}.

In Heckman's model, the outcome $Y_i$ is observed for individual $i$ if and only if $\tilde{Y}_i > 0$, which is equivalent to a probit selection model. The objective of inference is to estimate the parameter vector $\beta$. Heckman showed that
\begin{equation}
	\mathbb{E} (Y_i | X_i, \tilde{Y}_i > 0) = X_i^T \beta + \frac{\sigma_{12}}{\sigma_2} \lambda_i \label{heckman}
\end{equation}
where $\lambda_i$ is called the inverse Mills ratio and satisfies $\lambda_i = \frac{\phi(\Lambda_i)}{1 - \Phi(\Lambda_i)}$, $\Lambda_i = - \frac{\tilde{X}_i \gamma}{\sigma_2}$, using $\phi$, $\Phi$ to denote the density function and cumulative distribution function of a $N(0, 1)$ random variable respectively. The expectation $\mathbb{E} (Y_i | X_i, \tilde{Y}_i > 0)$ is taken over the observed data, while the inverse Mills ratio can be computed from the probit selection model. Therefore, Equation~\eqref{heckman} provides a way of estimating the full-data regression parameter $\beta$ using the observed data alone.

Heckman proposed a two-stage procedure in order to fit model \eqref{linear}-\eqref{heck.latent}. In the first step, probit regression is used to model the selection indicator $R = \{ \tilde{Y} > 0 \}$ in terms of fully observed covariates $\tilde{X}$. The fitted values from this regression are then used to estimate the inverse Mills ratios $\lambda_i$ for all individuals in the sample. In the second step, the outcome is regressed on the covariates $X_i$ and the inverse Mills ratios, using only data on individuals with observed outcome values. A parameter estimate for the regression coefficient $\beta$ can be obtained from this second-stage regression; we refer to Heckman's work for details on how to compute its standard error. Alternatively sample selection models can be fitted using maximum likelihood estimation \citep{Nawata1994}.

\subsection{The Method of Tchetgen Tchetgen and Wirth}

Heckman's sample selection model makes fairly restrictive parametric assumptions, requiring a normally distributed outcome and a probit model for the selection process. Various authors have attempted to relax these assumptions; we provide a brief literature review in the Supplementary Material. In this paper, we focus on an extension proposed by \cite{Tchetgen2017}. As previously, let $Z$ be an instrumental variable associated with the selection indicator $R$ but not with the outcome $Y$ conditional on observed covariates. To guarantee parameter identification across a wide range of statistical models, Tchetgen Tchetgen and Wirth impose the assumption that selection bias must be homogeneous on the scale of the parameter of interest. The exact mathematical form of the assumption depends on the model that is fitted; TTW considered linear, logistic and Poisson regression as examples. Here, we illustrate the assumption and estimation process for linear regression and delegate a discussion of logistic and Poisson regression to the Supplementary Material.


Suppose that a continuous outcome $Y$ is modelled using linear regression \eqref{linear}, as in Heckman's work. In this setting, selection bias can be quantified as the mean difference in outcome values between individuals with observed and unobserved outcomes:
\begin{equation}
	\mathbb{E} \left( Y | X, Z, R = 1 \right) - \mathbb{E} \left( Y | X, Z, R = 0 \right) = \delta (X, Z) \label{homog1}
\end{equation}
The homogeneous selection bias assumption states that this bias is only a function of the observed covariates and does not depend on the instrument: $\delta(X, Z) = \delta (X)$. Intuitively, this means that the instrument for selection should affect the chances of observing an individual participant's outcome but should not affect the overall magnitude of selection bias in the analysis. 

As demonstrated in Heckman's work, identification in linear regression is a consequence of the model assumptions and the homogeneous selection bias assumption is not necessary. However, the assumption can be used to derive a formula for estimation, similar to \eqref{heckman}:
\begin{equation}
	\mathbb{E} \left( Y | X, Z, R = 1 \right) = \mathbb{E} \left( Y | X \right) + \delta(X) (1 - \pi (X, Z)) \label{tch.lik1}
\end{equation}
where $\mathbb{E} (Y | X) = X^T \beta$ in the case of linear regression. The function $\pi (X, Z) = \mathbb{P} (R = 1 | X, Z)$ represents the propensity score for the selection indicator given $X$ and $Z$. Formula \eqref{tch.lik1} links the full-data regression $\mathbb{E} (Y | X)$ with the observed-data regression $\mathbb{E} \left( Y | X, Z, R = 1 \right)$ and can be used to construct a likelihood for $Y | X, Z, R = 1$. For linear regression with normally distributed errors, the likelihood has the form
\begin{equation*}
	\mathcal{L} (Y | X, Z, R = 1) = \prod_{i = 1}^n \left[ \left( \phi( Y_i ; X^T_i \beta + \delta(X_i) (1 - \pi (X_i, Z_i)), \sigma^2) \right)^{R_i} f_B (R_i; \pi(X_i, Z_i))  \right]
\end{equation*}
Here, $\phi (x; \mu, \sigma^2)$ is the normal density function and $f_B (x; \pi)$ is the Bernoulli probability mass function. If one is willing to make a set of parametric assumptions for the propensity score $\pi (X, Z)$ and the selection bias function $\delta (X)$, for example
\begin{equation*}
	\delta (X) = X^T \eta \;\;\; , \;\;\; \pi (X, Z) = (X \;\; Z)^T \alpha
\end{equation*}
then the likelihood can be maximized to yield consistent parameter estimates for the parameter vector $(\beta, \eta, \alpha, \sigma^2)$. Maximization can be implemented either as standard optimization over all the parameters simultaneously, or as a partial optimization procedure where one first maximizes the likelihood for the propensity score, $\prod_{i = 1}^n \left( f_B (R_i; \pi(X_i, Z_i)) \right)$, and then maximizes the likelihood for the outcome using the fitted values for the propensity score parameters. Although less flexible and less efficient in principle, the partial optimization procedure can have computational advantages in practice. Standard error estimates can be derived from the Hessian matrix at the optimal parameter values, as is common with maximum likelihood procedures. 


The TTW method is not restricted to linear regression and can be applied to more general models, either by relaxing the linearity assumption and considering non-linear (and possibly non-parametric) models where $\mathbb{E} (Y | X) = \mu(X; \beta)$, or by modelling non-normal outcomes e.g. via logistic or Poisson regression. The homogeneous selection bias assumption may not be strictly necessary for simple parametric models like linear regression, but becomes necessary for identification if the regression model is made flexible enough. We expand on these points in the Supplementary Material.

\section{Adjusting for Selection Bias in Mendelian Randomization}

\subsection{Missing Data in MR Studies}

We now discuss how to utilize ``instruments for selection" methods and adjust for selection bias due to MNAR data in Mendelian randomization studies. Some results in this direction have been reported in the econometrics literature \citep{Schwiebert2012}. We will use the abbreviation ``IVsel" to refer to the ``instruments for selection" methods, including both Heckman's sample selection and the TTW method.

In this section, we let $X$ denote the exposure and $Y$ the outcome of the MR analysis. Our objective is to infer the causal relationship between $X$ and $Y$, using a genetic instrument $G$ to account for the presence of unmeasured confounders $U$. We will assume that $G$ satisfies the traditional instrumental variable assumptions (is associated with $X$, is independent of $U$, and does not affect $Y$ by any other causal pathway except through $X$), and hence it is a valid instrument for inference in the absence of missing data.


Missing data can affect the MR analysis in three ways: data can be missing for the exposure but not the outcome, or for the outcome but not the exposure, or for both the exposure and the outcome. Each of these three scenarios can induce selection bias. For example, in an MR study with missing exposure values, if selection is only affected by the outcome $Y$, the data will be missing at random and selection bias will be eliminated by applying IPW. However, if the exposure $X$ also affects selection into the study, the data will be missing not at random and it will not be possible to fully model the selection process using the observed data alone.



\subsection{Instruments for Selection in MR}

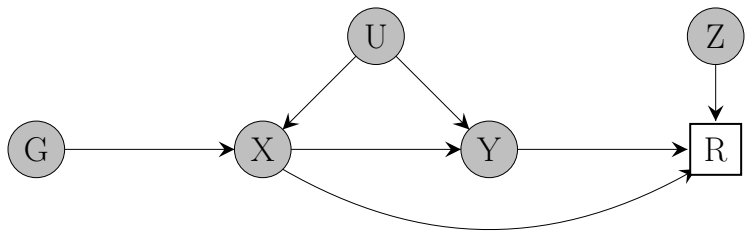
\begin{figure}
	\begin{center}
		\begin{tikzpicture}[scale=0.8, every node/.style={scale=0.8}]
		\draw [decoration={markings,mark=at position 1 with
			{\arrow[scale=2,>=stealth]{>}}},postaction={decorate}] (-3.5, 0) -- (-0.5, 0);
		\draw [decoration={markings,mark=at position 1 with
			{\arrow[scale=2,>=stealth]{>}}},postaction={decorate}] (1.65, 1.65) -- (0.35, 0.35);
		\draw [decoration={markings,mark=at position 1 with
			{\arrow[scale=2,>=stealth]{>}}},postaction={decorate}] (2.35, 1.65) -- (3.65, 0.35);
		\draw [decoration={markings,mark=at position 1 with
			{\arrow[scale=2,>=stealth]{>}}},postaction={decorate}] (0.5, 0) -- (3.5, 0);
		\draw [decoration={markings,mark=at position 1 with
			{\arrow[scale=2,>=stealth]{>}}},postaction={decorate}] (4.5, 0) -- (7.5, 0);
		\draw [decoration={markings,mark=at position 1 with
			{\arrow[scale=2,>=stealth]{>}}},postaction={decorate}] (8, 1.5) -- (8, 0.5);
		\draw [decoration={markings,mark=at position 1 with
			{\arrow[scale=2,>=stealth]{>}}},postaction={decorate}] (0.35, -0.35) to [out=330,in=210] (7.65, -0.35);
		\draw[fill = lightgray] (0, 0) circle (0.5cm);
		\draw[fill = lightgray] (-4, 0) circle (0.5cm);
		\draw[fill = lightgray] (4, 0) circle (0.5cm);
		\draw[fill = lightgray] (2, 2) circle (0.5cm);
		\draw[thick, fill = white] (7.55, 0.45) rectangle (8.45, -0.45);
		\draw[fill = lightgray] (8, 2) circle (0.5cm);
		\node (a) at (0, 0) {\Large{X}};
		\node (b) at (-4, 0) {\Large{G}};
		\node (c) at (4, 0) {\Large{Y}};
		\node (d) at (2, 2) {\Large{U}};
		\node (e) at (8, 0) {\Large{R}};
		\node (f) at (8, 2) {\Large{Z}};
		\end{tikzpicture}
	\end{center}
	\caption{A directed acyclic graph depicting the use of instruments for selection in Mendelian randomization. In this graph, both the exposure and the outcome are assumed to affect selection into the study.} 
\label{mr.dag}
\end{figure}

To account for selection bias, we leverage an instrumental variable $Z$ for the selection process, satisfying the IV assumptions of Section 2.1. The value of the instrument $Z$ must be observed for all individuals in the sample, including those with missing exposure and/or outcome measurements. Figure~\ref{mr.dag} illustrates how instruments for selection may be used to adjust for selection bias in MR studies.

The instrument for selection can be either genetic or non-genetic. Applications of Heckman's model in econometrics and social sciences have so far relied on non-genetic instruments. For example, when working with questionnaire data, potential instrumental variables can include interviewer characteristics (e.g. gender) or the mode of contact of study participants (by phone, email, etc). At the same time, when applying the IVsel methods in genetic epidemiology, it may be possible to obtain a genetic instrument for selection. This would require genetic information on both selected and unselected individuals to be available, so that genetic variants associated with selection (but not the exposure and/or outcome) can be identified. For example, \cite{Tyrell2021} identified genetic variants associated with participation in optional participation components of the UK Biobank dataset. These variants can be used as instruments in MR studies of exposures or outcomes from UK Biobank's optional participation components, in order to adjust for selection bias due to optional participation. As another example, in MR studies of disease progression, a genetic instrument for selection can be constructed using genetic variants associated with disease incidence but not with disease progression.

\subsection{MR with A Single Instrument for Inference}

We now discuss how the IVsel methods can be implemented in MR, when individual-level data are available.  First, we consider Mendelian randomization analyses using a single instrument for inference. This can be either a single genetic variant or an allele score comprising of multiple variants. Consider the simple MR model
\begin{eqnarray}
	X_i & = & \beta_X G_i + \epsilon_{Xi} \nonumber \\
	Y_i & = & \theta X_i + \epsilon_{Yi} \label{mr.model1} \\
	 & = & \theta \beta_X G_i + (\theta \epsilon_{Xi} + \epsilon_{Yi}) \nonumber \\
	 & = & \beta_Y G_i + \epsilon'_{Yi} \nonumber
\end{eqnarray}
with $(\epsilon_{Xi}, \epsilon_{Yi}) \sim N(0, \Sigma)$. The causal effect parameter $\theta$ is usually estimated by the Wald ratio,
\begin{equation}
	\hat{\theta} = \frac{\hat{\beta}_Y}{\hat{\beta}_X} \;\;\; , \;\;\; \text{s.e.} \left(\hat{\theta}\right) = \sqrt{\frac{\hat{s}_Y^2}{\hat{\beta}_X^2} + \frac{\hat{\beta}_Y^2 \hat{s}_X^2}{\hat{\beta}_X^4}} \label{wald}
\end{equation}
using a second-order approximation for the standard error. With individual data, the Wald ratio can be computed by regressing the exposure $X$ and outcome $Y$ respectively on $G$, and then taking the ratio of the regression coefficients from the two fits. The ordinary least squares (OLS) estimates $\hat{\beta}_X$, $\hat{\beta}_Y$ are consistent for $\beta_X$, $\beta_Y$, and provided that $\beta_X \neq 0$ (which is guaranteed by the IV assumptions), their ratio will be a consistent estimate of the causal effect $\theta$. This estimation strategy remains valid when the outcome $Y$ is binary \citep{Burgess2017review}.

With MNAR missing data for the exposure and/or outcome, the OLS estimates are biased. However, if an instrument for selection is available, Heckman's sample selection model or the TTW maximum likelihood procedure can be used to obtain asymptotically unbiased estimates of $\beta_X$, $\beta_Y$ and their standard errors. This needs to be done only for the variable for which there are missing data; for example, in a study where data are only missing for the outcome $Y$, the $G-X$ association can still be estimated by OLS. The estimates $\hat{\beta}_X, \hat{\beta}_Y$ can then be combined using the Wald ratio. The fact that this ratio is an asymptotically unbiased estimate of the $X-Y$ causal effect follows directly from standard MR theory. 

Regarding the IV assumptions for selection, the relevance assumption extends straightforwardly to MR studies and requires the instrument to be associated with selection into the study. The exclusion restriction assumption effectively requires the instrument to be independent of any variable for which data are missing. In an MR study with missing data only for the exposure, $Z$ must be independent of the exposure (conditional on the genetic instrument $G$ for inference) in order to adjust for selection bias when estimating the $G-X$ association. However, $Z$ can be associated with $Y$ without violating the exclusion restriction, since the $G-Y$ association can be estimated using OLS. In an MR study with missing outcome data, the instrument for selection must be independent of the outcome (conditional on $G$). In this case, a $Z-X$ association is also a violation of the exclusion restriction assumption, because, if $X$ is causal for $Y$, an association between the instrument and the exposure will open a pathway between $Z$ and $Y$ that is not mediated by $G$, and will induce bias in computing the $G-Y$ association estimate. Finally, in MR studies with missing data for both the exposure and the outcome, it follows trivially that the instrument for selection must be independent of both $X$ and $Y$ in the underlying population.

\subsection{MR with Multiple Instruments}

In practice, Mendelian randomization is often conducted using multiple instrumental variables for inference. A simple MR model with $K$ genetic instruments is
\begin{eqnarray}
	X_i & = & G_i^T \beta_X + \epsilon_{Xi} \nonumber \\
	Y_i & = & \theta X_i + \epsilon_{Yi} \label{multisnp.model} \\
	 & = & G_i^T \beta_Y + \epsilon'_{Yi} \nonumber \\
	(\epsilon_{Xi}, \epsilon_{Yi}) & \sim & N(0, \Sigma) \nonumber
\end{eqnarray}
where $G$ represents a $n \times K$ genetic data matrix and $\beta_X, \beta_Y$ are vector-valued. Common approaches for MR include two-stage least squares (2SLS) estimation when working with one-sample individual-level data, and the inverse variance weighted (IVW) formula when working with two-sample data. We will consider these two approaches separately and discuss how they can be combined with the IVsel methods.

\subsubsection{Two-Stage Least Squares}

The 2SLS procedure consists of first regressing the exposure $X$ on genetic variants $G_1, \dots, G_K$ and obtaining fitted exposure values $\hat{X}$, and then regressing the outcome $Y$ on the fitted values $\hat{X}$ and using the regression coefficient as an estimate of the causal effect $\theta$. The method requires access to one-sample individual-level data. Extensions of 2SLS can handle non-normal or non-linear exposure-outcome relationships by using fitted first-stage residuals instead of exposure values \citep{Terza2008}.

2SLS can be readily combined with IVsel approaches. With missing data for the exposure, the first-stage regression can be replaced by an implementation of the methods of Heckman or TTW. This will yield consistent $G-X$ association estimates, and hence fitted exposure values that are free of selection bias. The second-stage regression can be implemented as usual. With missing outcome data, instruments for selection can be used as part of the regression of $Y$ on $\hat{X}$ instead. In both cases, the causal effect estimates retain the consistency properties of the original 2SLS method. However, the same cannot be said about the estimated standard errors. In traditional 2SLS regression, the standard error associated with $\hat{\theta}_{2SLS}$ cannot be computed from the second-stage regression alone, as it needs to be adjusted for uncertainty in the first stage. The formula to compute the standard error is
\begin{eqnarray}
	\text{s.e.} \left( \hat{\theta}_{2SLS} \right) & = & \left( X^T G (G^T G)^{-1} G^T X \right)^{-1} \hat{\sigma}^2 \label{tsls.se} \\
	\hat{\sigma}^2 & = & \frac{1}{n - 1} (Y - X \theta)^T (Y - X \theta) \nonumber
\end{eqnarray}
This formula cannot be used in conjunction with the IVsel approaches. With missing data for the exposure, the vector $X$ is only partially observed and the standard error cannot be computed from \eqref{tsls.se}. On the other hand, with missing outcome data, the standard error \eqref{tsls.se} can be computed but the estimator to which it corresponds is biased. Instead, an unbiased estimate of the 2SLS standard error can be obtained using a bootstrap algorithm:
\begin{enumerate}
	\item Given data on individuals $i = 1, \dots, N$, repeat the following procedure $M$ times ($j = 1, \dots, M$):
	\begin{enumerate}
		\item Sample individuals $(i_1^{(j)}, \dots, i_N^{(j)}) \in \left\{ 1, \dots, N \right\}$ with replacement.
		\item For the sampled individuals, run 2SLS using Heckman's method or TTW to adjust for missing exposure or outcome data.
		\item Obtain a sample-specific causal effect estimate $\hat{\theta}^{(j)}$.
	\end{enumerate}
	\item Use the standard deviation of the $M$ values $\hat{\theta}^{(j)}$, $j = 1, \dots, M$ as an estimate of the 2SLS standard error. 
\end{enumerate}
However, this can be computationally intensive, especially when combined with the maximum likelihood procedure of the TTW method. With large numbers of genetic variants, a computationally efficient alternative could be to combine genetic variants into a single allele score and then use a Wald ratio estimate for the allele score.

\subsubsection{Selection-Adjusted Summary Statistics}

In practice, MR is often conducted using a two-sample framework in which genetic associations with the exposure and outcome are estimated in different datasets. This is particularly common when working with summary-level data. For genetic variant $G_j$, $j = 1, \dots, K$, let $\hat{\beta}_{Xj}$, $\hat{\beta}_{Yj}$ denote its association with the exposure and outcome respectively, and let $\hat{\sigma}_{Xj}^2$, $\hat{\sigma}_{Yj}^2$ denote the corresponding standard errors. The causal effect $\theta$ can be estimated from the inverse variance weighted formula,
\begin{equation}
	\hat{\theta}_{IVW} = \frac{\sum_j \hat{\beta}_{Xj} \hat{\beta}_{Yj} \hat{\sigma}_{Yj}^{-2}}{\sum_j \hat{\beta}_{Xj}^2 \hat{\sigma}_{Yj}^{-2}} \;\;\; , \;\;\; \text{s.e.} \left( \hat{\theta}_{IVW} \right) = \frac{1}{\sqrt{\sum_j \hat{\beta}_{Xj}^2 \hat{\sigma}_{Yj}^{-2}}} \label{ivw}
\end{equation}
With access to individual-level data, the summary statistics can be computed from a sequence of univariate linear regressions for each genetic variant separately, and then combined using IVW. In the presence of missing data, IVsel methods can be used to adjust the computation of summary statistics in a similar way as for MR with a single genetic instrument. The adjustment needs to be implemented for each genetic variant separately, meaning that the IVsel methods have to be implemented $K$ times, but the computational cost of doing so is not prohibitive. This procedure yields a set of ``selection-adjusted summary statistics" that can then be combined using the IVW formula \eqref{ivw}. Since estimates from the IVsel methods are consistent and asymptotically normal, this selection-adjusted IVW estimate will retain its standard theoretical properties.

The IVW estimator requires summary statistics $\hat{\beta}_{Xj}$, $\hat{\beta}_{Yj}$ to come from non-overlapping samples, therefore it is most commonly used in a two-sample MR setting. At the same time, the implementation of IVsel methods requires individual-level data. Therefore, the procedure described above is most efficient when having access to two separate individual-level datasets. However, gaining access to individual-level data from two separate sources can be difficult in practice. In the more common scenario of a single individual-level dataset, IVsel methods can still be combined with IVW but one has to guard against potential bias due to sample overlap. The effect of sample overlap is to exacerbate weak instrument bias \citep{Burgess2016bias}. Therefore, we would recommend that the IVsel methods are only combined with IVW in applications where the instruments for inference are strong and weak instrument bias is of lesser concern. Alternatively, the selection-adjusted summary statistics can be used in conjunction with recently proposed approaches to remove bias due to sample overlap \citep{Barry2021, Mounier2021}.

Individual-level data may not be necessary for both datasets in the two-sample setting. For example, in a two-sample MR study of disease progression, individual-level data will be needed for the disease progression trait in order to implement IVsel methods and adjust for selection bias due to conditioning on disease incidence. However, genetic associations with the exposure can be estimated from a summary-level dataset.

Access to individual-level data allows for estimating joint genetic effects, as well as marginal ones. This makes it straightforward to adjust for genetic correlations in MR analyses with correlated genetic variants. The two IVsel methods can be implemented with multiple covariates, meaning that modelling correlated variants remains straightforward in studies affected by sample selection.

Finally, recent years have seen the development of a plethora of MR methods using summary statistics, including adjustments for pleiotropic bias \citep[e.g.][]{Bowden2015, Bowden2016, Hartwig2017, Verbanck2018, Burgess2020conmix} and multivariable MR \citep{Sanderson2018, Zuber2020}, among other topics. In principle, selection-adjusted summary statistics can facilitate the use of such methods in applications with individual-level data and non-random missingness. However, this goes beyond the scope of the current paper.

\section{Simulation Study}

We implemented a simulation study to explore the performance of IVsel approaches in adjusting for selection bias. Simulations were conducted both for regression analyses, where the aim was to estimate the statistical association between a covariate $X$ and an outcome $Y$, and in a MR setting where the parameter of interest was the $X-Y$ causal effect. Although our focus in this paper is on applications to instrumental variable and MR analyses, we also conduct simulations for regression analyses. This is done for two reasons. First, the finite-sample performance of Heckman's sample selection model has been investigated in the past, but the relevant literature for the more recent TTW method has been limited. Second, MR estimates are derived as a combination of two regression-based estimates (e.g. the $G-X$ and $G-Y$ associations for a Wald ratio, or the two models fitted as part of a 2SLS algorithm), and adjusting for selection will affect MR estimates via these models. Thus it is important to understand how the IVsel methods perform in the regression framework in order to investigate their performance in MR analyses.

\subsection{Regression Analyses}

\subsubsection{Simulation Design}

First, we considered regression analyses. We assumed the existence of a single covariate $X$, whose association with the outcome $Y$ needs to be estimated. Keeping the notation consistent with Section 2, our baseline simulation setting was as follows:
\begin{eqnarray}
	Z, X, \epsilon & \sim & N(0, 1) \;\;\; \text{independently} \nonumber \\
	Y & = & \alpha + \beta X + \epsilon \nonumber \\
	R & \sim & \text{Logistic} (\pi_R) \nonumber \\
	\text{logit} \left( \pi_R \right) & = & \alpha_R + \beta_R X + \gamma_R Z + \delta_R Y \nonumber
\end{eqnarray}
The parameter of interest was the regression coefficient $\beta$. In our baseline simulation, we set $\beta = 0.1$, with this value chosen to facilitate a power comparison. The intercept $\alpha$ was set equal to 1. The parameter $\gamma_R$ determines instrument strength and was set to $\gamma_R = 0.4$ which corresponds to an $R^2$ statistic of $0.02$ for the instrument $Z$. The selection effects were set equal to $\beta_R = 0.5$ and $\delta_R = 0.5$, corresponding to an $R^2$ statistic of about $0.08$ for the covariate and outcome combined. Finally, the value of the intercept $\alpha_R$ was chosen to ensure that approximately $50\%$ of individuals had missing outcome data (in this baseline simulation, this meant that $\alpha_R = - \alpha \delta_R = - 0.5$). The simulation was implemented for $N = 10^4$ individuals.

We then considered the following modifications of our baseline scenario:
\begin{itemize}
    \item A null regression coefficient ($\beta = 0$).
	\item A binary instrument $Z$.
	\item A binary covariate $X$.
	\item A binary outcome (generated according to a logistic regression model).
	\item A discrete outcome (generated according to a Poisson regression model).
	\item Selection affected by the outcome but not the covariate.
	\item Selection affected by the covariate but not the outcome.
	\item A direct $Z-X$ effect.
	\item A direct (``pleiotropic") $Z-Y$ effect.
\end{itemize}
The direct $Z-X$ and $Z-Y$ effects were specified by modifying the generation of exposure and outcome values to be $X = \beta_X Z + \epsilon_X$, $\epsilon_X \sim N(0, 1)$ and $Y = \alpha + \beta X + \gamma Z + \epsilon$, $\epsilon \sim N(0, 1)$ respectively; for these simulations we set $\beta_X = \frac{1}{3}$ and $\gamma = \frac{1}{3}$, which correspond to an instrument $Z$ that explained approximately $10\%$ of variation in the exposure or outcome.

Along with our baseline scenario, there were a total of $10$ simulation settings. Each simulation was replicated $10^4$ times. 

In a second simulation, we explored how the performance of IVsel methods was affected by the strength of the instrument for selection. In our simulation, instrument strength is determined by the parameter $\gamma_R$. We varied the value of that parameter, setting it equal to $0$, $0.08$, $0.2$, $0.27$, $0.4$, $0.6$ and $0.95$. These values were selected to yield $R^2$ statistics of $0$, $0.1\%$, $0.5\%$, $1\%$, $2\%$, $5\%$ and $10\%$ respectively. Note that the $R^2$ statistic models the proportion of variation in the selection process that is explained by the instrument $Z$ in each simulation. All other aspects of the simulation design were kept the same as in our baseline scenario.

In each simulation scenario, we implemented Heckman's sample selection and the TTW method to estimate the regression coefficient $\beta$. We also implemented complete-case analysis (CCA), using only information on individuals with observed outcome values, as well as an inverse probability weighting (IPW) approach. These two methods rely on the MAR assumption and  are expected to yield biased results in our simulations. Finally, we also computed ``oracle" estimates using both observed and missing data, as a benchmark for comparison.

For the IPW method, we computed the inverse probability weights using a logistic regression model without interactions, which was the true model for $R$ in our simulation. We included the exposure $X$ but not the instrument $Z$ as explanatory variables in the logistic model; note that $Z$ is independent of the outcome, therefore its inclusion would offer little to no benefit to the IPW method \citep{Seaman2013}. The method was used simply as a benchmark for comparison with the IVsel methods in our simulations. As such, our implementation of IPW was quite simplistic; for example, we did not consider flexible weighting models and did not implement doubly robust estimation \citep{Kang2007}. These extensions can improve the performance of IPW in analyses with MAR data but would not be able to offer consistent estimation under our MNAR setting. For Heckman's method we included both $Z$ and $X$ in its selection model. For the TTW method, we used partial likelihood optimization; an implementation using the full likelihood yielded similar results. The method was implemented using a logistic model for the propensity score. It is possible to implement the method using a more flexible non-parametric propensity score model; indeed, this is one of the method’s advantages over Heckman’s approach. However, we have decided to use the logistic model here for simplicity. Standard errors were computed by inverting the (negative) Hessian matrix evaluated at the optimal parameter values and then taking the square roots of the diagonal elements, as is common with maximum likelihood-type methods. All simulations were conducted in the statistical software \texttt{R}. Heckman's method was implemented using the \texttt{R} package \texttt{sampleSelection} \citep{Toomet2008}. Standard errors for IPW were computed using the \texttt{survey} package \citep{Lumley2004}. The other methods were coded by hand. Our \texttt{R} code and simulation results are available at https://github.com/agkatzionis/IVsel.

\subsubsection{Simulation Results}

Simulation results are reported in Table~\ref{Obs.table1} for the 10 exploratory scenarios and in Table~\ref{Obs.table2} for the simulations with varying instrument strength. For each method, we report parameter estimates for the regression coefficient $\beta$, the empirical standard deviation of estimates across replications, model-based standard errors, the empirical coverage of $95\%$ confidence intervals and the empirical power to reject the null hypothesis $H_0: \; \beta = 0$ (except in the scenario where $\beta = 0$).

\begin{center}
\begin{table*}[t]%
\caption{Performance of selection bias adjustment methods in a range of different simulation scenarios.\label{Obs.table1}}
\centering
\scriptsize
\begin{tabular}{c@{\hskip 0.5in}ccccc@{\hskip 0.5in}ccccc}
  \hline
  \textbf{Method} & \textbf{Mean} & \textbf{Emp SD} & \textbf{StdErr} & \textbf{Cover} & \textbf{Power} & \textbf{Mean} & \textbf{Emp SD} & \textbf{StdErr} & \textbf{Cover} & \textbf{Power} \\ 
  \hline
   & \multicolumn{5}{c}{Baseline Scenario} & \multicolumn{5}{c}{$\beta = 0$} \\
  \hline
  CCA 		& 0.045 & 0.014 & 0.014 & 0.033 & 0.888 & -0.050 & 0.014 & 0.014 & 0.059 & ----- \\ 
  IPW 		& 0.045 & 0.011 & 0.011 & 0.001 & 0.984 & -0.050 & 0.010 & 0.011 & 0.003 & ----- \\ 
  Heckman 	& 0.099 & 0.024 & 0.024 & 0.952 & 0.989 &  0.000 & 0.022 & 0.022 & 0.950 & ----- \\ 
  TTW 		& 0.102 & 0.045 & 0.044 & 0.948 & 0.642 &  0.002 & 0.046 & 0.045 & 0.948 & ----- \\ 
  Oracle 	& 0.100 & 0.010 & 0.010 & 0.953 & 1.000 &  0.000 & 0.010 & 0.010 & 0.951 & ----- \\ 
  \hline
   & \multicolumn{5}{c}{Binary $Z$} & \multicolumn{5}{c}{Binary $X$} \\
  \hline
  CCA 		& 0.045 & 0.014 & 0.014 & 0.029 & 0.890 & -0.009 & 0.028 & 0.028 & 0.029 & 0.058 \\ 
  IPW		& 0.045 & 0.011 & 0.011 & 0.000 & 0.982 & -0.008 & 0.020 & 0.021 & 0.000 & 0.060 \\ 
  Heckman 	& 0.099 & 0.023 & 0.023 & 0.953 & 0.989 &  0.099 & 0.047 & 0.047 & 0.952 & 0.557 \\ 
  TTW 		& 0.102 & 0.045 & 0.044 & 0.946 & 0.639 &  0.110 & 0.174 & 0.175 & 0.951 & 0.098 \\ 
  Oracle 	& 0.100 & 0.010 & 0.010 & 0.949 & 1.000 &  0.100 & 0.020 & 0.020 & 0.949 & 0.999 \\ 
  \hline
   & \multicolumn{5}{c}{Binary Outcome} & \multicolumn{5}{c}{Discrete Outcome} \\
  \hline
  CCA 		& -0.009 & 0.038 & 0.038 & 0.176 & 0.058 &  0.071 & 0.008 & 0.008 & 0.071 & 1.000 \\ 
  IPW 		& -0.007 & 0.024 & 0.025 & 0.009 & 0.053 &  0.071 & 0.007 & 0.007 & 0.008 & 1.000 \\ 
  Heckman 	&  0.055 & 0.032 & 0.032 & 0.721 & 0.418 &  ----- & ----- & ----- & ----- & ----- \\ 
  TTW 		&  0.100 & 0.123 & 0.116 & 0.909 & 0.181 &  0.101 & 0.023 & 0.023 & 0.949 & 0.991 \\
  Oracle 	&  0.100 & 0.023 & 0.023 & 0.949 & 0.994 &  0.100 & 0.006 & 0.006 & 0.947 & 1.000 \\ 
  \hline
   & \multicolumn{5}{c}{No $X-R$ Effect} & \multicolumn{5}{c}{No $Y-R$ Effect (MAR)} \\
  \hline
  CCA 		& 0.089 & 0.013 & 0.013 & 0.878 & 1.000 &  0.100 & 0.015 & 0.015 & 0.948 & 1.000 \\ 
  IPW 		& 0.090 & 0.009 & 0.010 & 0.836 & 1.000 &  0.100 & 0.012 & 0.012 & 0.951 & 1.000 \\ 
  Heckman 	& 0.100 & 0.014 & 0.014 & 0.952 & 1.000 &  0.100 & 0.030 & 0.029 & 0.946 & 0.919 \\ 
  TTW 		& 0.101 & 0.076 & 0.075 & 0.946 & 0.274 &  0.099 & 0.042 & 0.041 & 0.948 & 0.675 \\ 
  Oracle 	& 0.100 & 0.010 & 0.010 & 0.952 & 1.000 &  0.100 & 0.010 & 0.010 & 0.951 & 1.000 \\ 
  \hline
   & \multicolumn{5}{c}{Added $Z-X$ Effect} & \multicolumn{5}{c}{Added $Z-Y$ Effect} \\
  \hline
  CCA 		& 0.035 & 0.014 & 0.014 & 0.004 & 0.727 &  0.029 & 0.015 & 0.015 & 0.002 & 0.497 \\ 
  IPW 		& 0.034 & 0.011 & 0.011 & 0.000 & 0.872 &  0.028 & 0.011 & 0.012 & 0.000 & 0.681 \\ 
  Heckman 	& 0.098 & 0.028 & 0.027 & 0.946 & 0.950 & -0.218 & 0.024 & 0.024 & 0.000 & 1.000 \\ 
  TTW 		& 0.103 & 0.041 & 0.040 & 0.947 & 0.721 & -0.183 & 0.042 & 0.039 & 0.000 & 0.995 \\ 
  Oracle 	& 0.100 & 0.010 & 0.009 & 0.949 & 1.000 &  0.100 & 0.011 & 0.011 & 0.951 & 1.000 \\ 
  \hline
\end{tabular}
\end{table*}
\end{center}

Our results for the baseline simulation scenario suggest that IVsel methods were able to adjust for selection bias due to missing data and returned unbiased estimates of the regression coefficient. The associated standard errors were fairly high compared to oracle estimates, resulting in low power to reject the null hypothesis $H_0: \; \beta = 0$, but the methods maintained nominal coverage and Type I error rates. Between the two methods, estimates from Heckman's method had smaller standard errors compared to the TTW method. IPW and complete-case analysis exhibited moderate degrees of bias, as expected. 

Similar results were obtained in the simulation with $\beta = 0$. Note that in this scenario, the coverage figures reported in Table~\ref{Obs.table1} equal one minus the Type I error rate, suggesting that IVsel methods attain nominal Type I error rates under a null regression coefficient.

For a binary instrument, IVsel methods were again able to adjust for selection bias while producing confidence intervals with nominal coverage. For a binary covariate, Heckman's method exhibited good performance but the results obtained from the TTW method were subject to very high uncertainty. For a binary outcome, Heckman's method exhibited a small degree of downward bias, while the TTW method obtained unbiased point estimates but exhibited slight undercoverage. In general, we have found the implementation of Heckman sample selection models in \texttt{R} to be less efficient for binary outcomes than for continuous ones. Nevertheless, both IVsel methods performed better than complete-case analysis and IPW. For a discrete outcome, the existing implementation of Heckman's method in \texttt{R} does not support fitting a Poisson model \citep{Terza1998}, therefore results from this method are not reported. The performance of the TTW method was satisfactory.

In simulations where selection was only a function of the outcome and not the covariate, the MAR methods remained biased but the bias was markedly smaller. The TTW approach resulted in higher standard errors compared to the baseline scenario, while Heckman's method produced smaller standard errors. This is further explored in Supplementary Material. The performance of all methods improved further in simulations where only the covariate, and not the outcome, affected selection. Since the covariate was fully observed in our simulations, this scenario represents a MAR mechanism, meaning that IPW can fully account for selection bias; indeed, the method exhibited no bias and nominal coverage in our simulations. Complete-case analysis was also unbiased, because the probability of observing the outcome was independent of the value of the outcome conditional on $X$ \citep{Hughes2019cca}. 

IVsel methods were able to accommodate an effect of the instrument on the covariate, as it does not violate the instrumental variable assumptions for $Z$. On the other hand, a direct $Z-Y$ effect constitutes a violation of the exclusion restriction assumption, and this induced bias in parameter estimates obtained from the IVsel methods. The magnitude of bias depends on the strength of the $Z-Y$ association, but our simulations indicate that the bias can be quite severe.

\begin{center}
\begin{table*}[t]%
\caption{Performance of selection bias adjustment methods in simulations with varying degrees of instrument strength for the instrument for selection.\label{Obs.table2}}
\centering
\scriptsize
\begin{tabular}{c@{\hskip 0.5in}ccccc@{\hskip 0.5in}ccccc}
  \hline
  \textbf{Method} & \textbf{Mean} & \textbf{Emp SD} & \textbf{StdErr} & \textbf{Cover} & \textbf{Power} & \textbf{Mean} & \textbf{Emp SD} & \textbf{StdErr} & \textbf{Cover} & \textbf{Power} \\ 
  \hline
   & \multicolumn{5}{c}{$R^2 = 0.1\%$} & \multicolumn{5}{c}{$R^2 = 0.5\%$} \\
  \hline
  CCA 		& 0.042 & 0.014 & 0.014 & 0.017 & 0.841 & 0.043 & 0.014 & 0.014 & 0.019 & 0.860 \\ 
  IPW 		& 0.042 & 0.011 & 0.011 & 0.001 & 0.968 & 0.043 & 0.011 & 0.011 & 0.001 & 0.973 \\ 
  Heckman 	& 0.094 & 0.090 & 0.090 & 0.966 & 0.167 & 0.099 & 0.040 & 0.040 & 0.956 & 0.703 \\ 
  TTW 		& 0.106 & 0.106 & 0.102 & 0.950 & 0.186 & 0.102 & 0.056 & 0.056 & 0.948 & 0.450 \\ 
  Oracle 	& 0.100 & 0.010 & 0.010 & 0.952 & 1.000 & 0.100 & 0.010 & 0.010 & 0.952 & 1.000 \\ 
  \hline
   & \multicolumn{5}{c}{$R^2 = 1\%$} & \multicolumn{5}{c}{$R^2 = 2\%$} \\
  \hline
  CCA 		& 0.044 & 0.014 & 0.014 & 0.021 & 0.874 & 0.045 & 0.014 & 0.014 & 0.033 & 0.888 \\ 
  IPW 		& 0.043 & 0.011 & 0.011 & 0.000 & 0.976 & 0.045 & 0.011 & 0.011 & 0.001 & 0.984 \\ 
  Heckman 	& 0.099 & 0.031 & 0.031 & 0.951 & 0.892 & 0.099 & 0.024 & 0.024 & 0.952 & 0.989 \\ 
  TTW 		& 0.103 & 0.051 & 0.050 & 0.949 & 0.543 & 0.102 & 0.045 & 0.044 & 0.948 & 0.642 \\ 
  Oracle 	& 0.100 & 0.010 & 0.010 & 0.949 & 1.000 & 0.100 & 0.010 & 0.010 & 0.953 & 1.000 \\ 
  \hline
   & \multicolumn{5}{c}{$R^2 = 5\%$} & \multicolumn{5}{c}{$R^2 = 10\%$} \\
  \hline
  CCA 		& 0.049 & 0.014 & 0.014 & 0.049 & 0.933 & 0.057 & 0.014 & 0.014 & 0.137 & 0.979 \\ 
  IPW 		& 0.049 & 0.011 & 0.011 & 0.002 & 0.994 & 0.056 & 0.011 & 0.011 & 0.018 & 1.000 \\ 
  Heckman 	& 0.100 & 0.018 & 0.019 & 0.955 & 0.999 & 0.100 & 0.016 & 0.016 & 0.952 & 1.000 \\ 
  TTW 		& 0.102 & 0.039 & 0.039 & 0.949 & 0.742 & 0.101 & 0.033 & 0.032 & 0.946 & 0.878 \\ 
  Oracle 	& 0.100 & 0.010 & 0.010 & 0.949 & 1.000 & 0.100 & 0.010 & 0.010 & 0.952 & 1.000 \\ 
  \hline
\end{tabular}
\end{table*}
\end{center}

Table~\ref{Obs.table2} contains results from our simulations exploring the performance of IVsel methods for varying degrees of instrument strength. The IVsel estimates were unbiased and attained nominal coverage regardless of the $R^2$ value, but weaker instruments led to more uncertainty about parameter estimates and lower power. The lack of bias is noteworthy: in MR analysis, using a weak instrument for inference will typically induce weak instrument bias in causal effect estimates. This happens because the MR estimand is effectively a ratio, and high uncertainty about the denominator of the ratio will not affect the value of the estimand in a symmetric way. On the other hand, in the simulations of Table~\ref{Obs.table2}, the estimand is a single parameter, and a weak instrument will produce a parameter estimate with high uncertainty but no bias.

Strong instruments resulted in IVsel estimates with high precision. When instrument strength was $5-10\%$, the accuracy of Heckman's sample selection method was comparable to that of IPW, while the TTW method exhibited more uncertainty than other approaches but had reasonable power. The increased uncertainty of the TTW method is not surprising, given that the method makes rather flexible modelling assumptions, but may also be a consequence of the misspecification of the propensity model $\pi (X, Z)$ (the IVsel methods assume a probit model but we used a logistic model in our simulations). Note that the TTW method is likely to be more sensitive to such misspecification than Heckman's algorithm, since correct specification of $\pi (X, Z)$ is necessary to establish identification.

Overall, this simulation suggests that although IVsel methods can yield valid inferences with weak instruments, it is still important to identify strong instruments when using the methods in practice.

\subsection{Mendelian Randomization}

\subsubsection{Simulation Design}

In our MR simulations, the aim was to estimate the causal effect of an exposure $X$ on an outcome $Y$ in the presence of missing (MNAR) data for the exposure and/or outcome. We implemented two sets of MR simulations, one using a single genetic instrument for inference and a second one using multiple instruments. For the simulations with a single instrument for inference, we used the following data-generating model:
\begin{eqnarray}
	G, Z, U, \epsilon_X, \epsilon_Y & \sim & N(0, 1) \nonumber \\
	X & = & \alpha_X + \beta_X G + \gamma_X U + \epsilon_X \nonumber \\	
	Y & = & \alpha_Y + \theta X + \gamma_Y U + \epsilon_Y \nonumber \\
	R & \sim & \text{Logistic} (\pi_R) \nonumber \\
	\text{logit} \left( \pi_R \right) & = & \alpha_R + \beta_R X + \gamma_R Z + \delta_R Y \nonumber
\end{eqnarray}
The genetic instrument $G$ was generated according to a normal distribution to represent an MR analysis where a polygenic risk score is used a an instrumental variable.

In simulations with multiple instruments for inference, we used $K = 10$ genetic variants. For each variant, we generated effect allele frequencies $f_j \sim U(0.1, 0.9)$ and allele counts $G_j \sim \text{Binomial} (2, f_j)$. We then generated exposure measurements as
\begin{equation*}
	X = \alpha_X + G^T \beta_X + \gamma_X U + \epsilon_X 
\end{equation*}
and then simulated the outcome and the selection indicator as previously described.

The effects of genetic instruments on the exposure were specified so that approximately $5\%$ of variation in $X$ was explained by $G$. This was achieved by setting $\beta_X = \sqrt{2 / 19}$ in simulations with a single instrument for inference, and generating the $\beta_{Xj}$ from a $N(0, 0.05^2)$ distribution, left-truncated at $0.15$, in simulations with multiple instruments. The causal effect of interest was assigned values $\theta = 0.2$ to facilitate a power comparison between the various methods, or  $\theta = 0$ to explore the methods' performance under the null causal hypothesis. The intercepts $\alpha_X, \alpha_Y$ were set equal to zero for simplicity, while the confounding effects $\gamma_X, \gamma_Y$ were set equal to 1.

For the selection effects $\beta_R, \delta_R$, we considered three scenarios. In our first scenario, selection was affected by the outcome but not the exposure ($\beta_R = 0, \delta_R = 1$). In the second scenario, selection was affected by the exposure but not the outcome ($\beta_R = 1, \delta_R = 0$). In the third scenario, both the exposure and the outcome were causes of selection ($\beta_R = \delta_R = 0.5$). In all scenarios, the $R^2$ statistic for the effects of $X$ and $Y$ on selection was approximately $20\%$. The instrument-selection effect was set equal to $\gamma_R = 0.5$, corresponding to an $R^2$ value of approximately $2\%$ for $Z$. Finally, the intercept $\alpha_R$ was again tuned to ensure that fully-observed data were available for $50\%$ of individuals.

We considered both one-sample and two-sample MR designs. With a single instrument for inference, we used the Wald ratio to estimate the causal effect in both cases. With multiple instruments, we used two-stage least squares for one-sample MR and the summary statistics approach for two-sample MR. The summary statistics approach can also be implemented in one-sample MR; simulation results from such an implementation are reported in Supplementary Table 5. 2SLS was implemented using the \texttt{R} package \texttt{ivreg}, with $100$ bootstrap iterations used to compute the standard errors for the IVsel methods. IVW was implemented using the \texttt{R} package \texttt{MendelianRandomization}. For Wald ratios and IVW estimates, we used second-order approximations for the MR standard errors. The sample size used in our simulations was $N = 10^4$ for one-sample MR and $N_1 = N_2 = 10^4$ for two-sample MR. The simulations were replicated $10^4$ times for MR with a single instrument for inference, and $10^3$ times for MR with multiple instruments, to reduce the computational burden.

\subsubsection{Simulation Results}

Between the value of the causal effect $\theta$, the selection effects $\beta_R, \delta_R$, the number of instruments for inference and the MR study design (one-sample or two-sample), our simulation resulted in 24 distinct scenarios. The results are reported in Table~\ref{MR.table1} for MR simulations with a single instrument for inference, and in Table~\ref{MR.table2} for simulations with multiple instruments. 

\begin{center}
\begin{table*}[t]%
\caption{Performance of the various selection bias adjustment methods in Mendelian randomization simulations with a single genetic instrument for inference.\label{MR.table1}}
\centering
\scriptsize
\begin{tabular}{c@{\hskip 0.5in}ccccc@{\hskip 0.5in}cccc}
  \hline
  \multirow{2}{*}{\textbf{Method}} & \multicolumn{5}{c}{\textbf{$\theta = 0.2$}} & \multicolumn{4}{c}{\textbf{$\theta = 0$}} \\
   & \textbf{Causal} & \textbf{Emp SD} & \textbf{StdErr} & \textbf{Cover} & \textbf{Power} & \textbf{Causal} & \textbf{Emp SD} & \textbf{StdErr} & \textbf{Type I} \\ 
  \hline
   & \multicolumn{9}{c}{One-sample MR \;\; - \;\; $Y \rightarrow R$} \\
  \hline
  CCA 		& 0.143 & 0.056 & 0.059 & 0.847 & 0.696 & -0.001 & 0.054 & 0.054 & 0.052 \\ 
  IPW 		& 0.143 & 0.039 & 0.049 & 0.847 & 0.886 & -0.001 & 0.038 & 0.044 & 0.023 \\ 
  Heckman 	& 0.199 & 0.061 & 0.065 & 0.962 & 0.882 & -0.002 & 0.058 & 0.058 & 0.053 \\ 
  TTW 		& 0.203 & 0.343 & 0.340 & 0.947 & 0.094 & -0.005 & 0.304 & 0.301 & 0.052 \\ 
  Oracle 	& 0.199 & 0.044 & 0.049 & 0.971 & 0.986 & -0.001 & 0.044 & 0.044 & 0.050 \\ 
  \hline
   & \multicolumn{9}{c}{One-sample MR \;\; - \;\; $X \rightarrow R$} \\
  \hline
  CCA 		& 0.266 & 0.060 & 0.068 & 0.866 & 0.985 & -0.001 & 0.059 & 0.058 & 0.051 \\ 
  IPW 		& 0.265 & 0.059 & 0.067 & 0.868 & 0.986 & -0.001 & 0.058 & 0.058 & 0.049 \\ 
  Heckman 	& 0.200 & 0.045 & 0.051 & 0.971 & 0.985 & -0.001 & 0.044 & 0.044 & 0.049 \\ 
  TTW 		& 0.202 & 0.075 & 0.073 & 0.957 & 0.814 & -0.001 & 0.046 & 0.046 & 0.053 \\ 
  Oracle 	& 0.199 & 0.044 & 0.049 & 0.972 & 0.987 & -0.001 & 0.044 & 0.044 & 0.050 \\ 
  \hline
   & \multicolumn{9}{c}{One-sample MR \;\; - \;\; $X, Y \rightarrow R$} \\
  \hline
  CCA 		& 0.054 & 0.070 & 0.072 & 0.484 & 0.112 & -0.119 & 0.068 & 0.064 & 0.449 \\ 
  IPW 		& 0.056 & 0.049 & 0.057 & 0.252 & 0.130 & -0.118 & 0.047 & 0.049 & 0.672 \\ 
  Heckman 	& 0.198 & 0.064 & 0.071 & 0.970 & 0.823 & -0.001 & 0.063 & 0.063 & 0.051 \\ 
  TTW 		& 0.182 & 0.421 & 0.387 & 0.970 & 0.036 & -0.029 & 0.974 & 0.919 & 0.004 \\ 
  Oracle 	& 0.199 & 0.044 & 0.049 & 0.974 & 0.988 & -0.001 & 0.044 & 0.044 & 0.049 \\    
  \hline
   & \multicolumn{9}{c}{Two-sample MR \;\; - \;\; $Y \rightarrow R$} \\
  \hline
  CCA 		& 0.144 & 0.058 & 0.059 & 0.841 & 0.690 &  0.001 & 0.053 & 0.054 & 0.048 \\ 
  IPW 		& 0.144 & 0.041 & 0.049 & 0.835 & 0.875 &  0.000 & 0.038 & 0.044 & 0.024 \\ 
  Heckman 	& 0.200 & 0.064 & 0.065 & 0.950 & 0.873 &  0.000 & 0.058 & 0.058 & 0.052 \\ 
  TTW 		& 0.213 & 0.344 & 0.340 & 0.946 & 0.101 & -0.003 & 0.305 & 0.301 & 0.053 \\ 
  Oracle 	& 0.200 & 0.049 & 0.049 & 0.951 & 0.984 &  0.000 & 0.044 & 0.044 & 0.054 \\ 
  \hline
   & \multicolumn{9}{c}{Two-sample MR \;\; - \;\; $X \rightarrow R$} \\
  \hline
  CCA 		& 0.268 & 0.068 & 0.068 & 0.831 & 0.980 &  0.000 & 0.059 & 0.058 & 0.054 \\ 
  IPW 		& 0.268 & 0.067 & 0.067 & 0.830 & 0.983 &  0.000 & 0.059 & 0.058 & 0.052 \\ 
  Heckman 	& 0.202 & 0.051 & 0.051 & 0.947 & 0.981 &  0.000 & 0.044 & 0.044 & 0.055 \\ 
  TTW 		& 0.204 & 0.081 & 0.074 & 0.950 & 0.791 &  0.000 & 0.046 & 0.046 & 0.055 \\ 
  Oracle 	& 0.201 & 0.050 & 0.049 & 0.946 & 0.984 &  0.000 & 0.044 & 0.044 & 0.053 \\ 
  \hline
   & \multicolumn{9}{c}{Two-sample MR \;\; - \;\; $X, Y \rightarrow R$} \\
  \hline
  CCA 		& 0.058 & 0.072 & 0.072 & 0.483 & 0.128 & -0.117 & 0.065 & 0.064 & 0.439 \\ 
  IPW 		& 0.057 & 0.051 & 0.057 & 0.264 & 0.149 & -0.117 & 0.046 & 0.049 & 0.668 \\ 
  Heckman 	& 0.202 & 0.071 & 0.072 & 0.949 & 0.806 &  0.000 & 0.063 & 0.063 & 0.051 \\ 
  TTW 		& 0.260 & 0.434 & 0.390 & 0.957 & 0.077 &  0.022 & 0.342 & 0.331 & 0.052 \\ 
  Oracle 	& 0.201 & 0.050 & 0.049 & 0.945 & 0.983 &  0.000 & 0.044 & 0.044 & 0.053 \\ 
  \hline
\end{tabular}
\end{table*}
\end{center}

In one-sample MR with only the outcome affecting missingness, the performance of IVsel methods was similar to that in regression analyses. The methods were able to adjust for selection bias at the expense of increased uncertainty around their causal effect estimates, which was not too pronounced for Heckman's method but was larger for TTW. The MAR approaches exhibited bias for $\theta = 0.2$ but not for $\theta = 0$; a null causal effect means that $G$ and $X$ are no longer upstream of the selection indicator in the causal diagram of Figure~\ref{mr.dag}, hence selection bias is not induced \citep{Gkatzionis2019}.

When missingness was affected by the exposure but not the outcome, the performance of IVsel methods improved further and the standard errors produced by the methods were not much larger than those of the oracle analysis. Recall that the standard error of the Wald ratio \eqref{wald} depends primarily on the standard error $\hat{s}_Y$ of the $G-Y$ association and only to a lesser extend on the standard error $\hat{s}_X$ of the $G-X$ association. When missingness is only a function of $X$, the IVsel methods will produce an increased standard error $\hat{s}_X$ but this will only have a small impact on $\text{s.e.} (\hat{\theta})$.

CCA and IPW were biased in this simulation, at least for $\theta \neq 0$, but the direction of bias was different to the scenario with missing outcome values. This is again due to the fact that the MR causal effect estimate is a ratio of two parameter estimates, $\hat{\theta} = \frac{\hat{\beta}_Y}{\hat{\beta}_X}$. In Section 4.1, CCA and IPW exhibited downward bias. If such bias arises in the numerator of the Wald ratio, it will result in an underestimation of the MR causal effect. On the other hand, downward bias in the denominator will result in overestimation. In the case of $\theta = 0$, selection bias did not appear with missing exposure data because all methods were able to estimate $\beta_Y = 0$ unbiasedly, meaning that any bias in estimating $\beta_X$ was of lesser concern.

When missingness was affected by both the exposure and the outcome, TTW estimates deviated slightly from the true causal effect values. This was because of the substantial uncertainty in $\beta_X$ estimates obtained from the TTW method, as seen in the previous section. Similar to how weak instrument bias occurs in traditional MR studies, when $\hat{\beta}_X$ takes a value close to zero, it can result in unstable causal effect estimates. This happened in a few replications of our simulation, due to the high uncertainty of the TTW method, hence giving the impression of bias. Heckman's method was less affected by this phenomenon because of the method's smaller standard errors.

Finally, a comparison of results from one-sample MR and two-sample MR simulations suggests little difference between the two settings in terms of the performance of selection bias adjustment methods.

\begin{center}
\begin{table*}[t]%
\caption{Performance of the various selection bias adjustment methods in Mendelian randomization simulations with multiple genetic instruments for inference.\label{MR.table2}}
\centering
\scriptsize
\begin{tabular}{c@{\hskip 0.5in}ccccc@{\hskip 0.5in}cccc}
  \hline
  \multirow{2}{*}{\textbf{Method}} & \multicolumn{5}{c}{\textbf{$\theta = 0.2$}} & \multicolumn{4}{c}{\textbf{$\theta = 0$}} \\
   & \textbf{Causal} & \textbf{Emp SD} & \textbf{StdErr} & \textbf{Cover} & \textbf{Power} & \textbf{Causal} & \textbf{Emp SD} & \textbf{StdErr} & \textbf{Type I} \\ 
  \hline
   & \multicolumn{9}{c}{One-sample MR (2SLS) \;\; - \;\; $Y \rightarrow R$} \\
  \hline
  CCA 		& 0.159 & 0.053 & 0.055 & 0.889 & 0.823 & 0.013 & 0.051 & 0.052 & 0.042 \\ 
  IPW 		& 0.152 & 0.040 & 0.046 & 0.872 & 0.940 & 0.006 & 0.037 & 0.043 & 0.028 \\ 
  Heckman 	& 0.205 & 0.059 & 0.060 & 0.955 & 0.938 & 0.010 & 0.054 & 0.056 & 0.048 \\ 
  TTW 		& 0.226 & 0.332 & 0.338 & 0.959 & 0.102 & 0.004 & 0.299 & 0.301 & 0.057 \\ 
  Oracle 	& 0.206 & 0.044 & 0.043 & 0.947 & 0.998 & 0.008 & 0.043 & 0.043 & 0.054 \\ 
  \hline
   & \multicolumn{9}{c}{One-sample MR (2SLS) \;\; - \;\; $X \rightarrow R$} \\
  \hline
  CCA 		& 0.066 & 0.084 & 0.083 & 0.638 & 0.126 & -0.138 & 0.084 & 0.083 & 0.363 \\ 
  IPW 		& 0.050 & 0.059 & 0.063 & 0.336 & 0.103 & -0.153 & 0.061 & 0.064 & 0.666 \\ 
  Heckman 	& 0.195 & 0.074 & 0.071 & 0.939 & 0.780 & -0.007 & 0.073 & 0.071 & 0.060 \\ 
  TTW 		& 0.130 & 0.046 & 0.042 & 0.588 & 0.852 &  0.006 & 0.034 & 0.033 & 0.062 \\ 
  Oracle 	& 0.208 & 0.042 & 0.043 & 0.951 & 0.999 &  0.007 & 0.043 & 0.043 & 0.049 \\ 
  \hline
   & \multicolumn{9}{c}{One-sample MR (2SLS) \;\; - \;\; $X, Y \rightarrow R$} \\
  \hline
  CCA 		& 0.077 & 0.066 & 0.067 & 0.541 & 0.203 & -0.103 & 0.065 & 0.065 & 0.333 \\ 
  IPW 		& 0.067 & 0.048 & 0.054 & 0.293 & 0.206 & -0.109 & 0.045 & 0.051 & 0.566 \\ 
  Heckman 	& 0.214 & 0.062 & 0.062 & 0.949 & 0.928 &  0.010 & 0.060 & 0.061 & 0.050 \\ 
  TTW 		& 0.379 & 0.219 & 0.218 & 0.876 & 0.425 &  0.187 & 0.221 & 0.198 & 0.186 \\ 
  Oracle 	& 0.208 & 0.043 & 0.043 & 0.947 & 0.997 &  0.007 & 0.042 & 0.043 & 0.047 \\ 
  \hline
   & \multicolumn{9}{c}{Two-sample MR (Summary Statistics) \;\; - \;\; $Y \rightarrow R$} \\
  \hline
  CCA 		& 0.141 & 0.058 & 0.062 & 0.853 & 0.642 &  0.002 & 0.053 & 0.057 & 0.032 \\ 
  IPW 		& 0.141 & 0.041 & 0.050 & 0.824 & 0.859 &  0.002 & 0.037 & 0.044 & 0.022 \\ 
  Heckman 	& 0.196 & 0.063 & 0.068 & 0.965 & 0.837 &  0.001 & 0.058 & 0.061 & 0.041 \\ 
  TTW 		& 0.220 & 0.349 & 0.368 & 0.965 & 0.078 & -0.008 & 0.316 & 0.321 & 0.041 \\ 
  Oracle 	& 0.195 & 0.049 & 0.052 & 0.958 & 0.968 &  0.001 & 0.044 & 0.046 & 0.037 \\ 
  \hline
   & \multicolumn{9}{c}{Two-sample MR (Summary Statistics) \;\; - \;\; $X \rightarrow R$} \\
  \hline
  CCA 		& 0.252 & 0.065 & 0.070 & 0.902 & 0.962 &  0.002 & 0.058 & 0.061 & 0.028 \\ 
  IPW 		& 0.254 & 0.064 & 0.070 & 0.898 & 0.970 &  0.002 & 0.058 & 0.061 & 0.026 \\ 
  Heckman 	& 0.190 & 0.049 & 0.052 & 0.960 & 0.965 &  0.002 & 0.043 & 0.046 & 0.027 \\ 
  TTW 		& 0.107 & 0.046 & 0.045 & 0.420 & 0.646 &  0.000 & 0.033 & 0.035 & 0.048 \\ 
  Oracle 	& 0.192 & 0.048 & 0.052 & 0.967 & 0.972 &  0.002 & 0.043 & 0.046 & 0.028 \\ 
  \hline
   & \multicolumn{9}{c}{Two-sample MR (Summary Statistics) \;\; - \;\; $X, Y \rightarrow R$} \\
  \hline
  CCA 		& 0.052 & 0.071 & 0.075 & 0.500 & 0.097 & -0.111 & 0.064 & 0.067 & 0.381 \\ 
  IPW 		& 0.056 & 0.051 & 0.057 & 0.261 & 0.139 & -0.115 & 0.045 & 0.051 & 0.643 \\ 
  Heckman 	& 0.193 & 0.073 & 0.070 & 0.931 & 0.783 &  0.001 & 0.063 & 0.063 & 0.054 \\ 
  TTW 		& 0.129 & 0.273 & 0.282 & 0.956 & 0.074 &  0.024 & 0.241 & 0.253 & 0.043 \\ 
  Oracle 	& 0.198 & 0.051 & 0.052 & 0.952 & 0.978 &  0.000 & 0.044 & 0.046 & 0.043 \\ 
  \hline
\end{tabular}
\end{table*}
\end{center}

Table~\ref{MR.table2} reports results from simulations using multiple genetic instruments for inference. A notable difference from simulations with a single instrument for inference was that the performance of the TTW method deteriorated. In two-sample MR, the method was still able to obtain unbiased estimates of the SNP-specific $G-X$ and $G-Y$ associations, but the precision of these estimates was quite low and combining these associations using the IVW formula produced biased causal effect estimates. Similar issues arose in the 2SLS implementation of the method. There are similarities here between the suboptimal performance of TTW estimates and MR analyses with measurement error, where $G-X$ and $G-Y$ associations are also estimated with low precision. Methods aiming to adjust MR analyses for measurement error \citep{Bowden2018} could potentially be used to improve the performance of the TTW method, but we have not explored this further.

The precision of TTW estimates depends on the sample size and the strength of the instrument for selection, meaning that the method may perform better in MR analyses with larger sample sizes and/or stronger instruments. In addition, a comparison of Tables~\ref{MR.table1} and \ref{MR.table2} suggests that TTW estimates were more accurate when the method only used a single instrument for inference. Therefore, it may be useful to combine genetic instruments into a polygenic risk score, if possible, and then use the TTW method with a single instrument for inference.

Heckman's method did not face the same issue as TTW due to the smaller degree of uncertainty about its estimates. The method produced estimates with little bias in all scenarios and was the preferred method in this simulation. The performance of complete-case analysis and IPW was similar to that observed in simulations with a single instrument for inference. Of note was the different direction of bias between one-sample and two-sample MR when selection was affected only by the exposure. This was due to the method used in each case. In one-sample MR, we used the 2SLS approach whose default implementation in \texttt{R} discards individuals with missing values from both the first and the second stage, meaning that selection bias will affect both the exposure and outcome models. The pattern of bias for 2SLS was therefore similar to the single-instrument simulations where selection was affected by both the exposure and outcome, and this also caused bias for $\theta = 0$. On the other hand, in two-sample MR we used IVW to obtain causal effect estimates. This requires summary statistics for the exposure and outcome to be estimated separately, and with no missing data in the sample for the outcome we were able to estimate unbiasedly the $G-Y$ associations. This resulted in a pattern of bias similar to that observed in simulations with a single instrument and selection depending only on $X$. In addition, bias in the two-sample case was not observed when $\theta = 0$.

Additional simulation scenarios, both for regression and for MR analyses, are considered in Supplementary Tables 1-6 and Supplementary Figures 2, 5 and 6. In these simulations, we vary some aspects of our simulation design and explore how these affect the performance of IVsel methods in regression and MR analyses. In particular, we consider simulations with different proportions of missing data, different strengths of the selection effects ($X-R$ and $Y-R$), violations of the IV assumptions for selection, model misspecification, MR analyses with different sample sizes, and different values of the MR causal effect.

\section{Application: Effects of BMI on Smoking in ALSPAC}

We implemented the selection bias adjustment methods in a real-data application, using data from the Avon Longitudinal Study of Parents and Children (ALSPAC) \citep{Boyd2013, Fraser2013}, a longitudinal population-based study that recruited pregnant women residing in Avon, UK, with expected delivery dates between 1st April 1991 and 31st December 1992. The study included $15454$ pregnancies resulting in $15589$ foetuses, $14901$ of which were alive at 1 year of age. Ethical approval for the study was obtained from the ALSPAC Ethics and Law Committee and the Local Research Ethics Committees. Informed consent for the use of data collected via questionnaires and clinics was obtained from participants following the recommendations of the ALSPAC Ethics and Law Committee at the time. The study website (http://www.bristol.ac.uk/alspac/researchers/our-data/) contains details on all data that is available through a fully searchable data dictionary and variable search tool.

Using data collected from the offspring, our objective in this application was to assess the relation between BMI and two smoking traits, namely smoking status (ever vs never) and the number of cigarettes smoked per day on average.

\subsection{Data and Methods}

Access to ALSPAC individual-level data was obtained under application B3838. We used data from both clinic visits (``Teen Focus 4" stage, age 18) and questionnaires (``It’s all about you", age 20). BMI measurements were recorded during the clinic visit. Outcome data for smoking status and number of cigarettes smoked per day were self-reported as part of the questionnaire. We also obtained genetic data for the MR analyses, and information on participants' sex and maternal variables (smoking before pregnancy, gestational age and highest educational qualification held), which were used as additional covariates in our analysis.

Missing data were present for all variables. Our aim was to investigate the performance of various methods in adjusting for missingness in the exposure and outcome, hence we restricted our analysis to genotyped individuals with fully recorded maternal covariates. This resulted in a sample size of $n = 7779$ individuals.

We performed Mendelian randomization to study the relationship between BMI and smoking outcomes. MR estimates were obtained as Wald ratios, using a polygenic risk score (PRS) as a single instrument for BMI. To construct the PRS, we obtained a list of genetic variants associated with BMI from a previous GWAS study \citep{Pulit2018}. In total, 311 variants were matched to our ALSPAC data. Variants were coded so that the effect alleles had risk-increasing effect on BMI. The score was then constructed using the SNP-BMI associations from the BMI GWAS as weights. The PRS-BMI association was estimated from linear regression. The association of the PRS with smoking status was estimated using logistic regression, while the association with the number of cigarettes smoked per day was estimated using linear regression, as well as Poisson regression.

\subsection{Instruments for Selection}

For the IVsel methods, we used four different instruments for selection. The first instrument was non-genetic. As part of the ALSPAC study, a small-scale randomized trial was conducted when inviting participants to complete the "It's all about you" questionnaire. Participants were randomized to either receive an online invitation, or to receive a printed version of the questionnaire and have a choice on whether to return it in print or online. Mode of delivery of the questionnaire affected its completion, albeit weakly \citep{Bray2017}. Not all ALSPAC participants were enrolled in the trial; consequently, our analyses using this instrument had to be conducted on a smaller set of $n_{RCT} = 5646$ individuals.

The other three instruments were genetic. A recent study investigated genetic associations with ALSPAC participation \citep{Taylor2018}, which can be used to construct a polygenic risk score for participation. In order to obtain a reasonably strong instrument, we used a lenient threshold of $10^{-4}$ for inclusion of genetic variants; this resulted in a risk score comprising $178$ variants, after LD pruning. More details on the construction of the PRS for participation are given in the Supplementary Material. 

To construct the last two instruments, we considered genetic associations with optional participation components in UK Biobank \citep{Tyrell2021}. We derived two polygenic risk scores, one for variants associated with completion of the Food Frequency questionnaire (FFQ) and one for the Mental Health Questionnaire (MHQ). These risk scores consisted of 13 and 40 SNPs respectively. Arguably, selection into UK Biobank is different to ALSPAC participation, but the two risk scores were reported to have high genetic correlations with ALSPAC participation \citep[$0.488$ for FFQ, $0.627$ for MHQ,][]{Tyrell2021}, and the SNPs included in these scores did not associate with BMI or any of the outcomes in our study, hence they constitute valid instruments.

\subsection{Results}

\begin{figure}[ht]
\begin{center}
	\includegraphics[scale = 0.45]{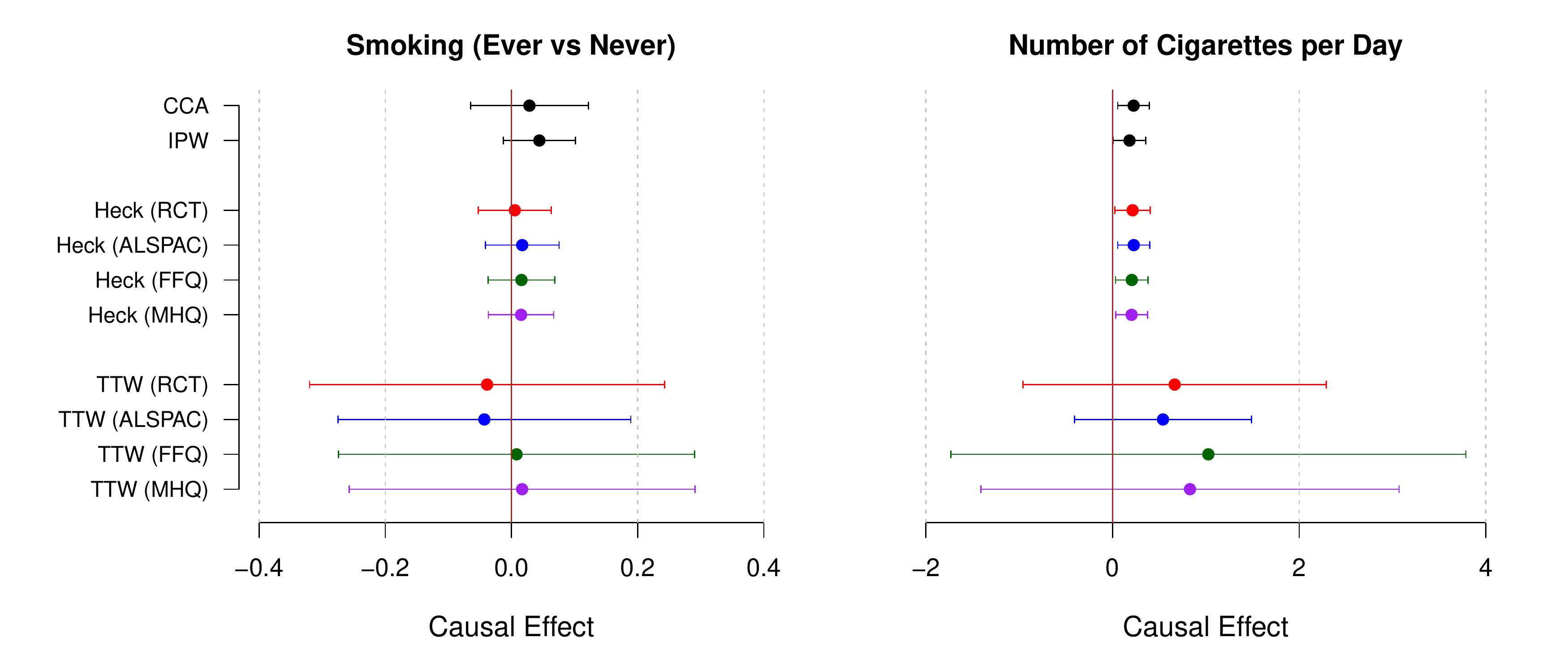}
\end{center}
\caption{MR estimates of the effect of BMI on smoking status (left) and number of cigarettes smoked per day (right), obtained from ALSPAC data using various methods to adjust for missing data in BMI and smoking outcomes. Colors indicate which instrument for selection is used each time (red: RCT for participation, blue: ALSPAC risk score, green: SNPs associated with FFQ completion in UK Biobank, purple: SNPs associated with MHQ completion in UK Biobank, black: method requires no instrument for selection).}
\label{alspac.fig}
\end{figure}

We implemented complete-case analysis, IPW and the IVsel methods to estimate the effects of BMI on the two smoking outcomes in our study. Results are reported in Figure~\ref{alspac.fig} and in Supplementary Table 7. Our analysis identified a risk-increasing effect of genetically elevated BMI on the number of cigarettes smoked per day, with effect estimates of $0.226$ ($95\%$ CI $(0.056, 0.396)$) by complete-case analysis and $0.182$ ($95\%$ CI $(0.007, 0.356)$) by inverse probability weighting. Heckman's method confirmed this result, with all four implementations of the method suggesting similar effect estimates as CCA and IPW. The TTW method suggested a larger effect, but the method's precision was low and the effect did not pass the $5\%$ significance threshold.

For smoking status, the point estimates and confidence intervals plotted in Figure~\ref{alspac.fig} represent log-odds ratios of increase in odds of smoking per unit increase in BMI. There was little evidence of a causal effect of genetically elevated BMI on smoking status in our analysis. Causal effect estimates were consistent across methods, although the TTW method again produced larger standard errors and wide confidence intervals.

Implementations of the IVsel methods using different instruments for selection produced similar causal effect estimates, which is reassuring. Among the four instruments, the ALSPAC-derived risk score was fairly strong, while the other three instruments were rather weak and produced F statistics near or below $10$ (Supplementary Table 10). Accordingly, the TTW method attained higher precision when implemented using the ALSPAC risk score, confirming the benefits of using strong instruments when applying the IVsel methods in practice. Results from Heckman's method exhibited less variation with different instruments; for binary outcomes, the method seemed to be less sensitive to instrument strength.

Results obtained using the IVsel approaches were in decent agreement with those obtained from the more traditional methods. Since the IVsel methods make more general assumptions about the missingness mechanism than CCA or IPW, their implementation here is useful as a form of sensitivity analysis, and suggests that selection bias is unlikely to have had a serious impact in our analysis.

\cite{Howe2017} used ALSPAC data to identify an effect of BMI on early-onset smoking (at age 15-16) in female participants, but no effect of BMI on late-onset smoking (age 17+) in either males or females. In this application we used smoking data from adulthood (age 20+), hence our findings are consistent with the lack of a late-onset effect in \cite{Howe2017}. In sex-stratified analyses, not reported here, we observed no effect of BMI on smoking status in either males males or females. Evidence of a reverse effect of smoking on BMI has been observed in \citep{Freathy2011}, meaning that early-onset smoking could act as a confounder between BMI and smoking at ages 18-20, which could induce bias in our analysis if BMI affects smoking at an early age. An exhaustive analysis of the relationship between BMI and smoking across different age groups \citep{Morris2021} would be a challenging task, hence we decided to limit our attention to the age groups in our dataset. We should mention, however, that studies using larger datasets, including the UK Biobank, have identified effects of obesity on smoking initiation \citep{Torres2018, Taylor2018smoking}, so our null result could be due to the small sample size of ALSPAC. This could also explain the seemingly counterintuitive finding that BMI associates with smoking intensity but not smoking status in our analysis: smoking status is a binary variable, hence the MR analysis with smoking status as an outcome has lower power than the analysis with number of cigarettes as an outcome.

\section{Discussion}

In this paper we have discussed how to use instruments for selection in order to adjust for selection bias due to MNAR data, in regression studies and MR analyses with individual-level data. This is an established approach in econometrics and has recently found applications in biomedical research \citep{Marra2017, Kone2019, Pirastu2021}. In this paper, we have argued that this approach can readily be extended to Mendelian randomization studies. We have conducted a simulation study to assess its performance, and utilized it in a real-data application studying the effects of BMI on smoking traits. 

There is an extensive literature on Heckman-type selection models and it would be difficult to cover it exhaustively. Instead, we have decided to focus on two approaches: Heckman's original method and its extension by Tchetgen Tchetgen and Wirth. Both methods were able to adjust for selection bias in a wide range of simulation scenarios, at the expense of lower precision in parameter estimates. The methods performed best with strong instruments and continuous covariates and outcomes. Potential violations of the instrumental variable assumptions should be carefully considered before using the methods in practice, as such violations can induce substantial biases. Comparing the two methods, Heckman's approach usually produced smaller standard errors; however, Heckman's normality assumptions mean that the TTW approach is more widely applicable, as illustrated in simulations with binary and discrete outcomes.

Our simulation study explored a wide range of scenarios and we hope it can provide guidelines to applied researchers about the performance of the IVsel methods in both traditional regression analyses and MR. Nevertheless it has not been possible to cover every simulation scenario of interest. For example, we did not vary the sample size and the number of genetic instruments for inference. A large sample size will increase the computational cost of running the IVsel methods but will also improve the methods' performance; on the other hand, a smaller sample size will further increase the uncertainty around causal effect estimates, which can affect TTW particularly. A large number of instruments for inference could be particularly worrying for the 2SLS approach because it will require solving a high-dimensional optimization problem to estimate the $G-X$ associations. We have also not considered simulations with interactions between variables in the selection model; such interactions are known to have a significant impact on the magnitude of selection bias \citep{Jiang2017, Shahar2017}.

Our work admits a number of methodological extensions. For example, it may be possible to formulate our IVsel-adapted 2SLS procedure as a structural equation model, which would improve its computational efficiency by facilitating estimation of standard errors and avoid the need to use a bootstrap algorithm. Another potential extension would be to use the sample selection framework in order to compute bounds for the MR causal effect under missing data. Such bounds have already been proposed for regression-based analyses, both in Tchetgen Tchetgen and Wirth's paper and elsewhere \citep{Lee2009, Marden2018, Honore2020}. The bounds reported in these papers can in principle be extended to MR; for example, when estimating a Wald ratio, one can compute bounds for the instrument-exposure and instrument-outcome associations separately, and combine them to derive worst-case bounds for the causal effect of interest. Note, however, that these bounds can only be computed for binary (or at least bounded) variables. In addition, the bounds can be very wide; for example, if the bounds for both the instrument-exposure and the instrument-outcome effect include $0$, the range of possible values for the causal effect of interest will span the entire real line. Another interesting extension of our work could be to combine the IVsel approaches with methods that offer robustness to weak instruments. Such methods may include limited-information maximum likelihood for point estimation and Anderson-Rubin or conditional likelihood ratio tests for hypothesis testing. Using these approaches in conjunction with the IVsel methods may improve the methods' performance, especially for the TTW method. In addition, it is worth noting that Heckman-type selection models are not the only approach for addressing selection bias due to MNAR data. For example, some authors have explored the use of shadow variables instead of instruments for selection \citep{Miao2015}. Another approach is to use multiple imputation, for example through the NARFCS algorithm \citep{Tompsett2018}. This approach has already proved useful in traditional epidemiologic studies, but to the best of our knowledge it has not been implemented in MR studies. Some attempts to combine imputation with Heckman-type selection models have also been made in the literature \citep{Galimard2018}. On the other hand, the literature on two-sample summary-data MR has also given rise to selection bias adjustment methods, designed specifically for studies of disease progression. Such methods include \cite{Dudbridge2019}'s index event bias adjustment and SlopeHunter \citep{Mahmoud2020}. A comparison between IVsel and these methods is beyond the scope of the current paper, but would be an interesting direction of future research.

Some \texttt{R} code to implement the IVsel methods is available in the GitHub page \\
https://github.com/agkatzionis/IVsel. We hope that our code and results will prove useful to applied researchers looking to address issues related to selection bias and missing data in their work.


\section*{Acknowledgments}

The authors are grateful to Dr Joshua Bell and Dr Hannah Jones for their assistance in creating polygenic risk scores for BMI and ALSPAC participation respectively. Apostolos Gkatzionis and Kate Tilling received funding for this project by the UK Medical Research Council and the University of Bristol (MRC-IEU core funding, MC UU 00011/3). Jon Heron was also supported by this grant. Eric Tchetgen Tchetgen received support from grants by the U.S. National Institute of Health.

Access to ALSPAC data was obtained as part of application B3838. The UK Medical Research Council and Wellcome (Grant ref: 217065/Z/19/Z) and the University of Bristol provide core support for ALSPAC. A comprehensive list of grants funding the ALSPAC study is available on the ALSPAC website (http://www.bristol.ac.uk/alspac/external/documents/grant-acknowledgements.pdf). GWAS data was generated by Sample Logistics and Genotyping Facilities at Wellcome Sanger Institute and LabCorp (Laboratory Corporation of America) using support from 23andMe. We are extremely grateful to all the families who took part in this study, the midwives for their help in recruiting them, and the whole ALSPAC team, which includes interviewers, computer and laboratory technicians, clerical workers, research scientists, volunteers, managers, receptionists and nurses.





\subsection*{Conflict of interest}

The authors declare no potential conflict of interests.

\section*{Supporting information}

The following supporting information is available as part of the online article:

\noindent
\textbf{Supplementary File S1.}
{Supporting information for the paper, including a brief summary of the TTW method for logistic and Poisson regression, additional simulation results and more details about the real-data application.}

\noindent
\textbf{Supplementary Data S2.}
{Genetic variants, effect alleles and weights used to create the polygenic score for BMI, as well as the scores used as instruments for ALSPAC participation.}

\bibliography{ReFusion}
\bibliographystyle{chicago}

\end{document}


\title{\textbf{\Large{Using Instruments for Selection to Adjust for Selection Bias in Mendelian Randomization}}}
\author{Apostolos Gkatzionis$^{1,2}$, Eric J. Tchetgen Tchetgen$^{3}$, Jon Heron$^{1,2}$, \\
Kate Northstone$^{2}$, Kate Tilling$^{1,2}$}
\date{\begin{small}$^1$MRC Integrative Epidemiology Unit, University of Bristol, UK.
$^2$Population Health Science Institute, Bristol Medical School, University of Bristol, UK.
$^3$Department of Statistics and Data Science, Wharton School, University of Pennsylvania, USA.
\end{small}}
\maketitle

\vspace{1.5cm}

\begin{center}
\textbf{\Large{Supplementary Material}}
\end{center}

This part contains the supplementary material for the paper ``Using Instruments for Selection to Adjust for Selection Bias in Mendelian Randomization". It includes a brief summary of the TTW method for logistic and Poisson regression, additional simulation results, as well as more details about our real-data application.

\vspace{0.8cm}

\section*{Relaxing Heckman's Modelling Assumptions}

An important limitation of Heckman's sample selection model is that it requires a normally distributed outcome and a probit model for the selection process. Various extensions or modifications of Heckman's modelling assumptions have been proposed in the literature. Parametric extensions include a probit model for binary outcomes \citep{Vandeven1981}, a Poisson distribution for discrete outcomes \citep{Terza1998} and a $t$ distribution for continuous outcomes \citep{Marchenko2012} among others. Many of these extensions relied on maximum likelihood estimation, instead of Heckman's original two-stage algorithm, which is harder to extend to non-normal settings.

In terms of non-parametric extensions, \cite{Newey1990} and \cite{Ahn1993} used Heckman's two-stage approach but replaced the normal regression and probit selection models with semi- or non-parametric equivalents. \cite{Das2003} combined Heckman's probit selection model with a non-parametric model for the outcome, $Y = \mu (X) + \epsilon$, and used splines to model the regression function $\mu (X)$. One of the main challenges these non-parametric approaches faced was to establish parameter identification. This is not a concern in many parametric models, including Heckman's original setting, because the restrictive nature of a parametric model is often sufficient to guarantee identification. However, it can be a hard task in non-parametric settings; for example, Das et al. established identification for their basis function expansion algorithm only up to an additive constant.

As discussed in our manuscript, \cite{Tchetgen2017} proposed general-purposed general-purpose identification condition and developed a maximum likelihood estimation algorithm that can be applied to a wide range of parametric models. We presented the algorithm for linear regression in the main part of our manuscript, and refer to the next Section of this Supplement for an overview of logistic and Poisson regression. The appendix of Tchetgen Tchetgen and Wirth's paper also contains a comparison of their work with that of \cite{Das2003}. In related work, \cite{Sun2018} proposed an alternative identification condition, which can be easier to verify than the homogeneous selection bias assumption, along with a different estimation strategy.

Finally, some authors have considered using the sample selection framework in order to compute bounds for the MR causal effect under missing data \citep{Lee2009, Marden2018, Honore2020}.

\section*{TTW for Logistic and Poisson Regression}

In this section, we outline the TTW method for binary and discrete outcomes. We focus on stating the homogeneous selection bias assumption and outlining the estimation process, and refer to \cite{Tchetgen2017} for more details.

\subsection*{Logistic Regression}

Suppose that the outcome $Y$ is a binary variable, and that we wish to model it using logistic regression. In this case, our interest lies in estimating the log-odds ratio
\begin{equation*}
	\text{logit} \mathbb{P} (Y = 1 | X) = \log \frac{\mathbb{P} (Y = 1 | X)}{\mathbb{P} (Y = 0 | X)} = X^T \beta
\end{equation*}
Selection bias in the odds ratio scale can be quantified as the log-odds ratio for $Y = 1$ between individuals with observed and unobserved outcome values:
\begin{equation}
	\log \frac{\mathbb{P} (Y = 1 | X, Z, R = 1) / \mathbb{P} (Y = 0 | X, Z, R = 1)}{\mathbb{P} (Y = 1 | X, Z, R = 0) / \mathbb{P} (Y = 0 | X, Z, R = 0} = \omega (X, Z) \label{homog2}
\end{equation}
The homogeneous selection bias assumption states that selection bias must be independent of the instrument on a log-odds ratio scale, so $\omega(X, Z)$ must be a function of $X$ only and cannot depend on $Z$. As with linear regression, this assumption is not strictly necessary for identification in logistic regression models but can be used to establish identification in more general models where the goal is to make inference on the regression function $\mu (X; beta) = \text{logit} \mathbb{P} (Y | X)$.

Under the homogeneous selection bias assumption, the full-data regression parameter $\beta$ of a logistic regression model can be estimated using observed data via the relation
\begin{equation}
	\text{logit} \mathbb{P} (Y = 1 | X, Z, R = 1) = X^T \beta + \omega (X) - \log \left( e^{\omega(X)} \lambda (X, Z) + 1 - \lambda (X, Z) \right) \label{tch.lik2}
\end{equation}
where $\lambda (X, Z) = \mathbb{P} (R = 1 | X, Z, Y = 0)$ satisfies
\begin{equation}
	\pi (X, Z) = \frac{1}{1 + e^{X^T \beta}} \lambda(X, Z) + \frac{e^{X^T \beta}}{1 + e^{X^T \beta}} \frac{1}{1 + (1 - \lambda(X, Z)) e^{- \omega(X)} / \lambda(X, Z)} \label{tch.lik2.1}
\end{equation}
Equations~\eqref{tch.lik2}-\eqref{tch.lik2.1} can be used to construct a likelihood for the observed-data regression and the propensity score respectively. Subject to parametric assumptions for $\omega(X)$ and $\lambda(X, Z)$, the likelihood can then be optimized to estimate the regression coefficient $\beta$, similar to the linear regression case that we described in the main part of our manuscript. One difference is that, because the selection function and linear predictor term are included both in \eqref{tch.lik2} and in \eqref{tch.lik2.1}, it is not possible to split the maximum likelihood procedure in two parts and implement partial-likelihood optimization, therefore full optimization over all parameters simultaneously is the only option.

\subsection*{Poisson Regression}

If the outcome $Y$ represents count data, a common modelling assumption is to assume that they come from a Poisson distribution. Written as a generalized linear model, Poisson regression admits the logarithmic link function:
\begin{equation*}
	\mathbb{E} (Y | X) = \exp \left\{ X^T \beta \right\}
\end{equation*}
The homogeneous selection bias assumption again functions on the same scale as the regression coefficient of interest. Selection bias can be quantified as the ratio of outcome values between individuals with observed and unobserved outcomes,
\begin{equation}
	\frac{\mathbb{E} (Y | X, Z, R = 1)}{\mathbb{E} (Y | X, Z, R = 0)} = \nu (X, Z) \label{homog3}
\end{equation}
and the homogeneous selection bias assumption states that $\nu (X, Z)$ must be a function of $X$ but not $Z$. Similar to linear and logistic regression, the assumption can be used to obtain a relationship between the full-data regression coefficient $\beta$ and the observed-data regression,
\begin{equation}
	\log \mathbb{E} (Y | X, Z, R = 1) = X^T \beta + \log \nu (X) - \log \left( \nu (X) \pi (X, Z) + 1 - \pi (X, Z) \right) \label{tch.lik3}
\end{equation}
This formula facilitates a maximum likelihood estimation process for $\beta$ that only requires access to the observed data, along with parametric assumptions for the selection function $\nu (X)$ and the propensity score $\pi (X, Z)$. The same homogeneity assumption and maximum likelihood estimation procedure can be used for extensions of the standard Poisson regression model where $\mu(X) = \log \mathbb{E} (Y | X)$ is not assumed to have a linear form. 

\subsection*{Assessing the Homogeneous Selection Bias Assumption}

The homogeneous selection bias assumption offers a general, unifying framework for establishing parameter identification for the TTW method in a wide range of statistical models. However, the assumption can be difficult to verify in practice, since the selection function ($\delta$, $\omega$ or $\nu$) will depend on both observed and unobserved outcome values.

In their work, Tchetgen Tchetgen and Wirth identified a class of statistical models where the homogeneity assumption can be shown to hold. This class contains models in which the outcome $Y$ does not directly influence the selection indicator $R$, but instead there are common causes $U$ that influence both the outcome and selection, and the instrument $Z$ does not interact with $U$ in the selection model. The causal diagram in Supplementary Figure~\ref{Supp.dag1} provides an illustration. The homogeneous selection bias assumption is more likely to hold if effect modifiers for the $Z-R$ association are included as covariates in the $X-Y$ regression, which is hence recommended by the authors. 

In applications not characterized by Supplementary Figure~\ref{Supp.dag1}, intuitive arguments can be made about the validity of the homogeneous selection bias assumption. For example, when using questionnaire data, it has been suggested to use interviewer characteristics as an instrumental variable. Such interviewer characteristics can affect an individual's likelihood of responding to questions, but any effect they may have on the overall magnitude of selection bias is likely to be small, so the homogeneity assumption can be claimed to hold approximately.

We mention that \cite{Sun2018} proposed an alternative identification and modelling strategy for non-parametric estimation of statistical models with outcome data missing not at random. Their identification assumption can be easier to verify in practice, while estimation relies on extensions of established techniques such as IPW and outcome regression. The scope of that paper is hence similar to the TTW method presented here, but we have not explored it further and refer to the paper for more details.







\section*{Additional Simulation Results}

We now present results from additional simulations that complement those reported in the paper. As in the main part of our paper, we first discuss simulations for regression analyses and then consider simulations for MR studies.

\subsection*{Regression Analyses}

In the regression setting, we conducted six additional sets of simulations. First, we replicated the simulations of Table 1 with a null association between the covariate and the outcome ($\beta = 0$). Second, we considered different values for the selection effect parameters ($\alpha_R, \beta_R, \gamma_R, \delta_R$). Third, we induced $Z-X$ and $Z-Y$ associations of varying strength and studied how these affect the simulation results. Fourth, we briefly considered a weak-instrument simulation where the instrument was also binary. Fifth, we considered a simulation scenario in which the outcome $Y$ is not a cause of selection $R$, but the two variables are confounded instead. And sixth, we conducted simulations in which the outcome was continuous but not normally distributed, to asses how this form of model misspecification affects the performance of the various methods.

\subsubsection*{Null $X-Y$ Association}

First, we provide simulation results for the scenarios considered in Table 1 of the main part of the paper, but implemented for a null value of the parameter of interest. We ran these simulations by setting $\beta = 0$ instead of $\beta = 0.1$, and keeping all other aspects of the simulation design unchanged. In Table~\ref{Supp.table1} below, we report average parameter estimates, their empirical standard deviation, model-based standard errors and the Type I error rate of $95\%$ confidence intervals for these simulations.

\begin{center}
\begin{table*}[t]%
\caption{Performance of the various selection bias adjustment methods in regression analyses, for various simulation scenarios and $\beta = 0$.\label{Supp.table1}}
\centering
\scriptsize
\begin{tabular}{c@{\hskip 0.5in}cccc@{\hskip 0.5in}cccc}
  \hline
  \textbf{Method} & \textbf{Mean} & \textbf{Emp SD} & \textbf{StdErr} & \textbf{Type I} & \textbf{Mean} & \textbf{Emp SD} & \textbf{StdErr} & \textbf{Type I} \\ 
  \hline
   & \multicolumn{4}{c}{Binary $Z$} & \multicolumn{4}{c}{Binary $X$} \\
  \hline
  CCA 		& -0.050 & 0.014 & 0.014 & 0.942 & -0.103 & 0.028 & 0.028 & 0.951 \\ 
  IPW 		& -0.050 & 0.011 & 0.011 & 0.997 & -0.103 & 0.020 & 0.021 & 0.999 \\ 
  Heckman 	&  0.000 & 0.022 & 0.022 & 0.050 &  0.000 & 0.046 & 0.046 & 0.047 \\ 
  TTW 		&  0.002 & 0.046 & 0.046 & 0.050 &  0.008 & 0.175 & 0.175 & 0.051 \\ 
  Oracle 	&  0.000 & 0.010 & 0.010 & 0.050 &  0.000 & 0.020 & 0.020 & 0.050 \\ 
  \hline
   & \multicolumn{4}{c}{Binary Outcome} & \multicolumn{4}{c}{Discrete Outcome} \\
  \hline
  CCA 		& -0.016 & 0.006 & 0.006 & 0.831 & -0.025 & 0.008 & 0.008 & 0.866 \\ 
  IPW 		& -0.104 & 0.024 & 0.024 & 0.990 & -0.026 & 0.006 & 0.007 & 0.975 \\ 
  Heckman 	& -0.002 & 0.032 & 0.032 & 0.056 &  ----- & ----- & ----- & ----- \\ 
  TTW 		&  0.006 & 0.119 & 0.112 & 0.083 &  0.001 & 0.024 & 0.025 & 0.046 \\ 
  Oracle 	&  0.000 & 0.004 & 0.004 & 0.050 &  0.000 & 0.006 & 0.006 & 0.051 \\   
  \hline
   & \multicolumn{4}{c}{No $X-R$ Effect} & \multicolumn{4}{c}{No $Y-R$ Effect (MAR)} \\
  \hline
  CCA 		&  0.000 & 0.013 & 0.013 & 0.048 &  0.000 & 0.015 & 0.015 & 0.051 \\ 
  IPW 		&  0.000 & 0.009 & 0.010 & 0.036 & -0.000 & 0.012 & 0.012 & 0.051 \\ 
  Heckman 	&  0.000 & 0.014 & 0.014 & 0.048 &  0.000 & 0.029 & 0.029 & 0.046 \\ 
  TTW 		&  0.000 & 0.079 & 0.077 & 0.054 &  0.000 & 0.041 & 0.041 & 0.050 \\ 
  Oracle 	&  0.000 & 0.010 & 0.010 & 0.047 &  0.000 & 0.010 & 0.010 & 0.049 \\ 
  \hline
   & \multicolumn{4}{c}{Added $Z-X$ Effect} & \multicolumn{4}{c}{Added $Z-Y$ Effect} \\
  \hline
  CCA 		& -0.061 & 0.014 & 0.014 & 0.995 & -0.065 & 0.015 & 0.015 & 0.993 \\ 
  IPW 		& -0.061 & 0.011 & 0.011 & 1.000 & -0.065 & 0.011 & 0.011 & 1.000 \\ 
  Heckman 	& -0.001 & 0.026 & 0.026 & 0.051 & -0.289 & 0.023 & 0.023 & 1.000 \\ 
  TTW 		&  0.002 & 0.041 & 0.041 & 0.053 & -0.257 & 0.044 & 0.041 & 1.000 \\ 
  Oracle 	&  0.000 & 0.009 & 0.009 & 0.050 &  0.000 & 0.011 & 0.011 & 0.052 \\
  \hline
\end{tabular}
\end{table*}
\end{center}

The pattern of selection bias of this simulation was very similar to what we observed for $\beta = 0.1$. This suggests that the true value of the $X-Y$ association has little impact on the magnitude of selection bias in regression analyses. The IVsel methods produced unbiased point estimates and maintained nominal Type I error rates in all simulation scenarios except in the presence of pleiotropy ($Z \rightarrow Y$). Of note was the fact that Heckman's method exhibited a reasonable performance in the simulation with a binary outcome.

\subsubsection*{Selection Effects}

In our next simulation experiment, we varied the values of the selection effect parameters $\alpha_R, \beta_R, \gamma_R, \delta_R$. Recall that $\alpha_R$ represents the intercept in a logistic regression model for the selection indicator $R$, while $\beta_R, \gamma_R, \delta_R$ represent the effects of the covariate, instrument and outcome on the selection indicator respectively. Since $\gamma_R$ represents instrument strength, we have already explored the performance of IVsel methods for various values of that parameter (Table 2 in the main part of the paper), but the results are presented again here for completeness.

\begin{figure}[h]
\begin{center}
	\includegraphics[scale = 0.67]{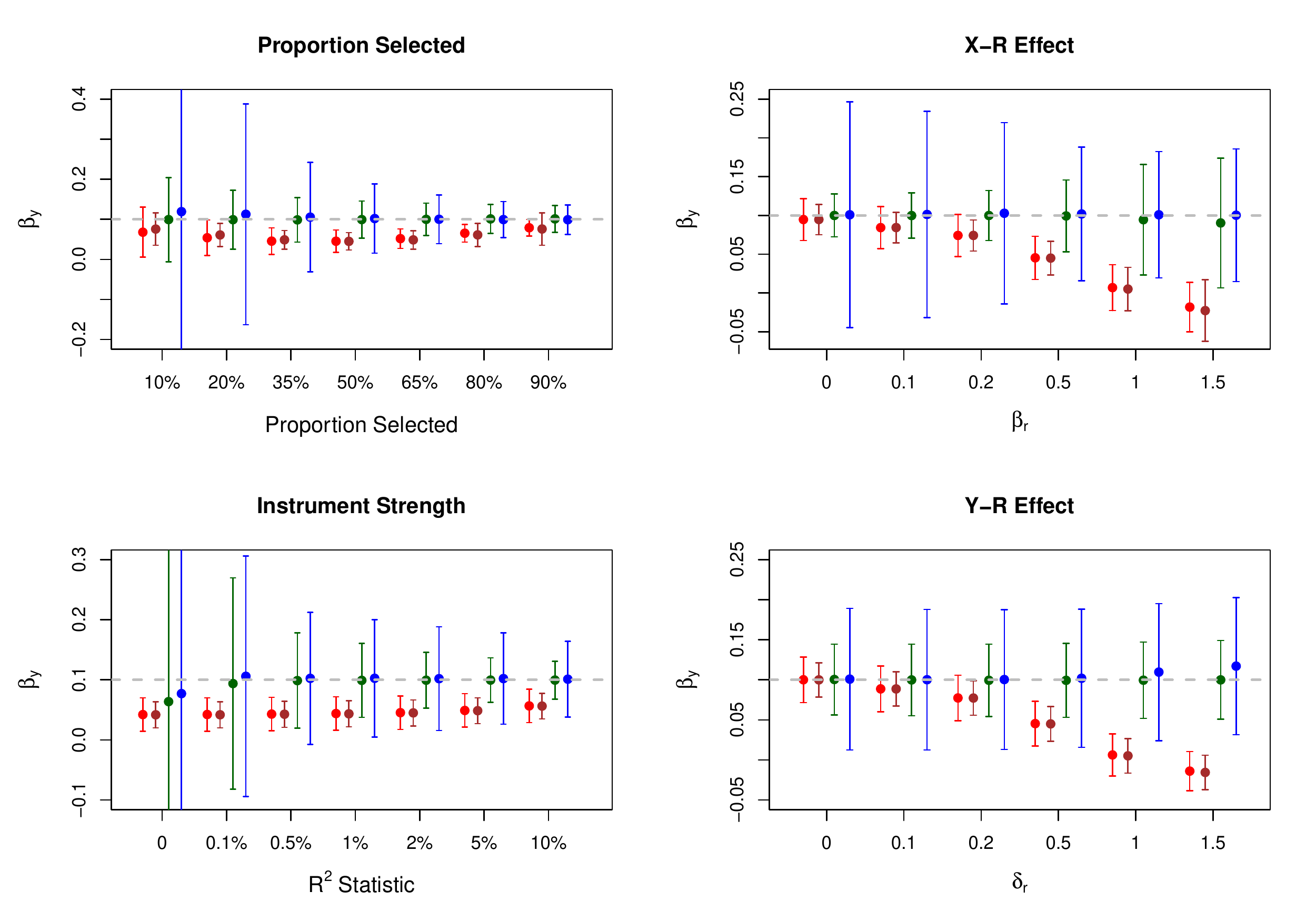}
\end{center}
\caption{Performance of the various selection bias adjustment methods in regression analyses, for different values of the selection effect parameters $\alpha_R$ (top left), $\beta_R$ (top right), $\gamma_R$ (bottom left) and $\delta_R$ (bottom right).}
\label{Supp.figure1}
\end{figure}

We varied the values of these four parameters one at a time, starting from the baseline simulation scenario in which $\alpha_R = - 0.5$, $\beta_R = 0.5$, $\gamma_R = 0.4$, $\delta_R = 0.5$. The intercept $\alpha_R$ determines the proportion of individuals for which the outcome is observed; in our baseline simulation, that proportion was $50\%$. In the new simulations, we used $\alpha_R$ values that resulted in $10\%$, $20\%$, $35\%$, $65\%$, $80\%$ and $90\%$ of individuals having observed outcomes. For the effects of the covariate ($\beta_R$) and outcome ($\delta_R$) on selection, we experimented with the values $0$, $0.1$, $0.2$, $0.5$, $1$, $1.5$. Note that the simulations with $\beta_R = 0$ or $\delta_R = 0$ correspond to scenarios where only one of $X, Y$ affects selection, similar to those reported in Table 1 in the main part of the paper. Finally, for the instrument strength parameter $\gamma_R$, we used the values $0.08$, $0.2$, $0.27$, $0.4$, $0.6$ and $0.95$ which corresponded approximately to $R^2$ statistic values of $0.1\%$, $0.5\%$, $1\%$, $2\%$, $5\%$ and $10\%$ respectively, as previously discussed. All simulations were conducted for a target parameter of $\beta = 0.1$. For ease of presentation, the results are reported in graphical form in Figure~\ref{Supp.figure1}. 

The IVsel methods remained unbiased regardless of the proportion of individuals for which the outcome was observed, while for IPW and complete case analysis, selection bias was stronger when $50\%$ of individuals were selected. This value results in the strongest $X-Y$ interaction in a log-probability model for selection, which is known to increase the effects of selection bias \citep{Jiang2017, Shahar2017}. The standard errors of all methods were larger for smaller values of $\alpha_R$, due to the fact that the $X-Y$ association parameter had to be estimated in a smaller sample. This was particularly problematic for the TTW method, whose standard errors were too wide to be of practical use when the outcome was observed in only $10\%$ or $20\%$ of individuals. 

Differences in the performance of the various methods were observed when we varied the strength of the $X-R$ association. The MAR-based methods exhibited small bias for small values of $\beta_R$. Larger values of $\beta_R$ meant stronger selection effects and bigger differences between individuals for which the outcome is observed compared to those for which the outcome is unobserved, so the impact of selection bias on complete-case and IPW estimates increased. Heckman's method remained unbiased for strong $X-R$ associations, but exhibited increased standard errors. Surprisingly, the opposite pattern was observed for the TTW method, whose performance improved for large $\beta_R$ values. We speculate that this may be due to the fact that a stronger $X-R$ association means the method has more information with which to model the selection process.

As discussed in the main part of our manuscript, the instrument strength parameter $\gamma_R$ affects the precision of IVsel estimates. The IVsel methods remain unbiased and maintain nominal coverage across $\gamma_R$ values, but the uncertainty about their parameter estimates increases substantially for weak instruments, at least when the methods are implemented for continuous outcomes. Figure~\ref{Supp.figure1} confirms these observations. The performance of IPW and complete-case analysis is barely affected by $\gamma_R$, since these two methods did not utilize the instrument $Z$ (the slight differences observed in the performance of CCA and IPW are because modifying the proportion of variation in selection that is explained by the instrument $Z$ will also modify the proportion of variation explained by $X$ and $Y$ and slightly alter the performance of these methods).

Finally, we performed simulations with different values of the $Y-R$ selection effect. The two MAR methods were unbiased for $\delta_R = 0$, since this simulation corresponds to a MAR setting, and exhibited increasing degrees of bias for larger values of $\delta_R$. The performance of IVsel methods was less affected by the value of $\delta_R$, although a slight overestimation of the regression parameter was observed for the TTW method for large $\delta_R$ values.

\subsubsection*{Z-X and Z-Y Effects}

In the main part of our manuscript, we conducted two simulations where we added a direct effect of the instrument for selection $Z$ on the exposure or outcome, and investigated how it affects the performance of the IVsel methods. Our conclusion was that an instrument-exposure effect is not problematic, but an instrument-outcome effect violates the second instrumental variable assumption for $Z$ and induces bias.

As an extension of these simulations, in this subsection we varied the strength of the $Z-X$ and $Z-Y$ effects. For $Z-X$ effects, we modified the baseline simulation design by generating exposure values as $X = \beta_X Z + \epsilon_X$, $\epsilon_X \sim N(0, 1)$. The values we used for $\beta_X$ were $0, 0.1, 0.2, 0.3, 0.5, 1$, with $\beta_X = 0$ corresponding to the baseline simulation from the main part of our paper. For $Z-Y$ effects, we aimed to explore the magnitude of bias induced under mild violation of the second instrumental variable assumption (a weak-to-moderate instrument-outcome effect). We therefore generated outcome values as $Y = \alpha + \beta X + \gamma Z + \epsilon$, letting the parameter $\gamma$ take values $0, 0.1, 0.2, 0.3, 0.5, 1$ (again, $\gamma = 0$ corresponds to our baseline simulation). Other aspects of the simulation design were left unchanged.

\begin{figure}[h]
\begin{center}
	\includegraphics[scale = 0.67]{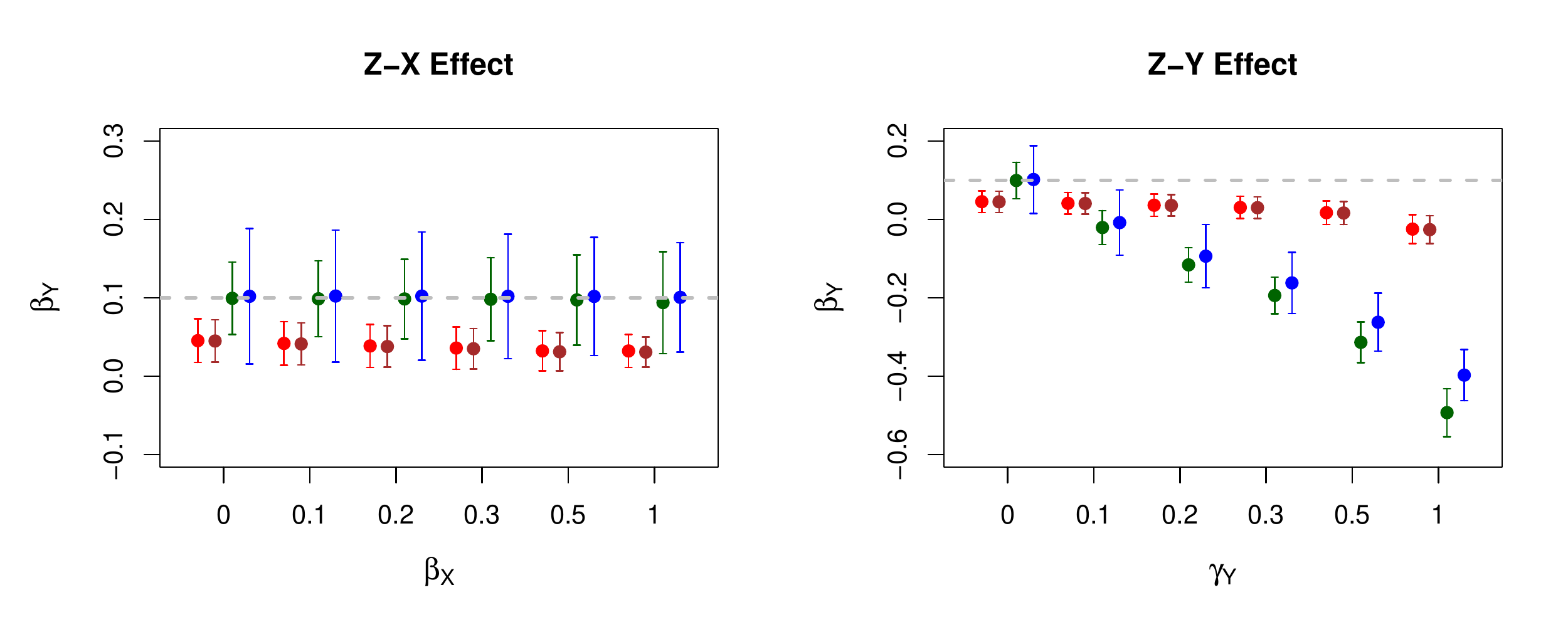}
\end{center}
\caption{Performance of the various selection bias adjustment methods in regression analyses with an added instrument-exposure (left) or instrument-outcome effect (right).}
\label{Supp.figure1.1}
\end{figure}

The results of this set of simulations are reported in Figure~\ref{Supp.figure1.1}. They confirm that the IVsel methods are robust to the presence of instrument-exposure effects; for all values of the parameter $\beta_X$, the methods gave similar results. On the other hand, even mild violations of the second instrumental variable assumption were enough to cause bias. Even for the relatively small value of $\gamma = 0.1$, point estimates from Heckman's method and TTW were further away from the true value on average than complete-case and IPW estimates. This suggests that care must be taken when selecting the instrument $Z$ to ensure it is independent of the outcome conditional on exposure.

\subsubsection*{Weak Binary Instruments}

In our next set of simulations, we explored whether the impact of weak instrument bias changes when using a binary instrument, instead of a continuous one. We did so by considering a single simulation scenario, similar to the "binary instrument" scenario we considered in the main part of the paper, but with the instrument strength parameter set to $\gamma_R = 0.23$, resulting in a simulation where the instrument explained approximately $0.2\%$ of variation in the selection process. This simulation was implemented for either a null ($\beta = 0$) or a positive ($\beta = 0.1$) parameter value. Other aspects of the simulation design were left unchanged. 

\begin{center}
\begin{table*}[t]%
\caption{Performance of the various selection bias adjustment methods in regression analyses with a weak binary instrument for selection.\label{Supp.table2}}
\centering
\scriptsize
\begin{tabular}{c@{\hskip 0.5in}ccccc@{\hskip 0.5in}cccc}
  \hline
  \multirow{2}{*}{\textbf{Method}} & \multicolumn{5}{c}{\textbf{$\beta = 0.1$}} & \multicolumn{4}{c}{\textbf{$\beta = 0$}} \\
 & \textbf{Mean} & \textbf{Emp SD} & \textbf{StdErr} & \textbf{Cover} & \textbf{Power} & \textbf{Mean} & \textbf{Emp SD} & \textbf{StdErr} & \textbf{Type I} \\ 
  \hline
  CCA 		&  0.042 & 0.014 & 0.014 & 0.018 & 0.848 & -0.053 & 0.014 & 0.014 & 0.962 \\ 
  IPW 		&  0.042 & 0.011 & 0.011 & 0.000 & 0.969 & -0.053 & 0.011 & 0.011 & 0.999 \\ 
  Heckman 	&  0.097 & 0.065 & 0.066 & 0.960 & 0.315 & -0.003 & 0.061 & 0.062 & 0.038 \\ 
  TTW 		&  0.104 & 0.078 & 0.077 & 0.948 & 0.273 &  0.002 & 0.076 & 0.075 & 0.048 \\ 
  Oracle 	&  0.100 & 0.010 & 0.010 & 0.950 & 1.000 &  0.000 & 0.010 & 0.010 & 0.050 \\ 
  \hline
\end{tabular}
\end{table*}
\end{center}

Table~\ref{Supp.table2} reports the results of this simulation experiment. Once again, IVsel methods remained unbiased despite the low instrument strength. Comparing these results with the binary-instrument simulations in Table~\ref{Supp.table1} and in the main part of the paper, we can also confirm that the precision of IVsel estimates decreased due to low instrument strength. These observations are similar to those obtained when exploring the performance of IVsel methods with weak continuous instruments; there was little change in our results when working with a binary instrument instead of a continuous one.

\subsubsection*{Outcome-Selection Confounding}

\begin{figure}
	\begin{center}
		\begin{tikzpicture}[scale=0.8, every node/.style={scale=0.8}]
		\draw [decoration={markings,mark=at position 1 with
			{\arrow[scale=2,>=stealth]{>}}},postaction={decorate}] (0.5, 0) -- (3.5, 0);
		\draw [decoration={markings,mark=at position 1 with
			{\arrow[scale=2,>=stealth]{>}}},postaction={decorate}] (9.65, 1.65) -- (8.35, 0.35);
		\draw [decoration={markings,mark=at position 1 with
			{\arrow[scale=2,>=stealth]{>}}},postaction={decorate}] (0.35, -0.35) to [out=330,in=210] (7.65, -0.35);
		\draw [decoration={markings,mark=at position 1 with
			{\arrow[scale=2,>=stealth]{>}}},postaction={decorate}] (5.65, 1.65) -- (4.35, 0.35);
		\draw [decoration={markings,mark=at position 1 with
			{\arrow[scale=2,>=stealth]{>}}},postaction={decorate}] (6.35, 1.65) -- (7.65, 0.35);
		\draw[fill = lightgray] (0, 0) circle (0.5cm);
		\draw[fill = lightgray] (4, 0) circle (0.5cm);
		\draw[thick, fill = white] (7.55, 0.45) rectangle (8.45, -0.45);
		\draw[fill = lightgray] (10, 2) circle (0.5cm);
		\draw[fill = lightgray] (6, 2) circle (0.5cm);
		\node (a) at (0, 0) {\Large{X}};
		\node (c) at (4, 0) {\Large{Y}};
		\node (e) at (8, 0) {\Large{R}};
		\node (f) at (10, 2) {\Large{Z}};
		\node (g) at (6, 2) {\Large{V}};
		\end{tikzpicture}
	\end{center}
\caption{A directed acyclic graph depicting the scenario where the outcome does not cause selection, but a confounder $V$ affects both the outcome and selection.}
\label{Supp.dag1}
\end{figure}
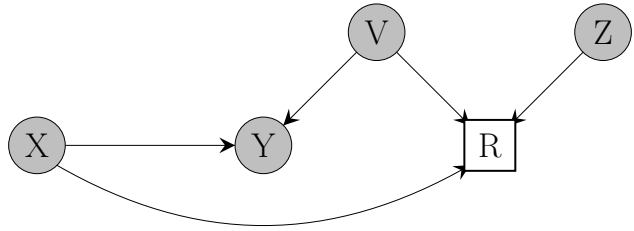

In our third set of simulations, we considered a scenario in which the outcome $Y$ is not a direct cause of the selection indicator $R$, but instead there is an unobserved confounding variable $V$ that affects both the outcome and the selection indicator. This scenario is depicted in the directed acyclic graph of Figure~\ref{Supp.dag1}. 

There are two reasons why this scenario is interesting. First, it was considered in the appendix of Tchetgen Tchetgen and Wirth's paper as a case where the homogeneous selection bias assumption is more likely to hold. The authors prove that the homogeneity assumption will be satisfied if $V$ is a continuous variable, $V$ does not interact with $Z$ in the selection model, and the outcome $Y$ is independent of $R$ and $Z$ conditional on $V, X$. We refer to the paper for a more formal statement of this result.

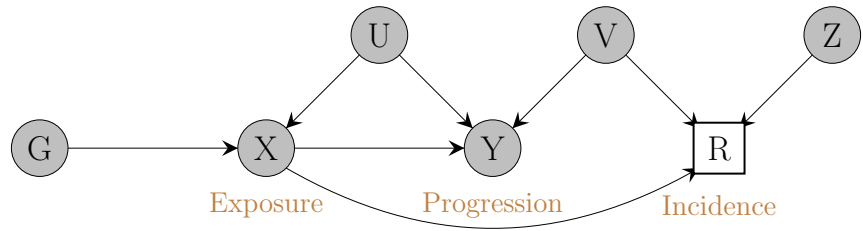
\begin{figure}
	\begin{center}
		\begin{tikzpicture}[scale=0.8, every node/.style={scale=0.8}]
		\draw [decoration={markings,mark=at position 1 with
			{\arrow[scale=2,>=stealth]{>}}},postaction={decorate}] (-3.5, 0) -- (-0.5, 0);
		\draw [decoration={markings,mark=at position 1 with
			{\arrow[scale=2,>=stealth]{>}}},postaction={decorate}] (1.65, 1.65) -- (0.35, 0.35);
		\draw [decoration={markings,mark=at position 1 with
			{\arrow[scale=2,>=stealth]{>}}},postaction={decorate}] (2.35, 1.65) -- (3.65, 0.35);
		\draw [decoration={markings,mark=at position 1 with
			{\arrow[scale=2,>=stealth]{>}}},postaction={decorate}] (0.5, 0) -- (3.5, 0);
		\draw [decoration={markings,mark=at position 1 with
			{\arrow[scale=2,>=stealth]{>}}},postaction={decorate}] (9.65, 1.65) -- (8.35, 0.35);
		\draw [decoration={markings,mark=at position 1 with
			{\arrow[scale=2,>=stealth]{>}}},postaction={decorate}] (0.35, -0.35) to [out=330,in=210] (7.65, -0.35);
		\draw [decoration={markings,mark=at position 1 with
			{\arrow[scale=2,>=stealth]{>}}},postaction={decorate}] (5.65, 1.65) -- (4.35, 0.35);
		\draw [decoration={markings,mark=at position 1 with
			{\arrow[scale=2,>=stealth]{>}}},postaction={decorate}] (6.35, 1.65) -- (7.65, 0.35);
		\draw[fill = lightgray] (0, 0) circle (0.5cm);
		\draw[fill = lightgray] (-4, 0) circle (0.5cm);
		\draw[fill = lightgray] (4, 0) circle (0.5cm);
		\draw[fill = lightgray] (2, 2) circle (0.5cm);
		\draw[thick, fill = white] (7.55, 0.45) rectangle (8.45, -0.45);
		\draw[fill = lightgray] (10, 2) circle (0.5cm);
		\draw[fill = lightgray] (6, 2) circle (0.5cm);
		\node (a) at (0, 0) {\Large{X}};
		\node (b) at (-4, 0) {\Large{G}};
		\node (c) at (4, 0) {\Large{Y}};
		\node (d) at (2, 2) {\Large{U}};
		\node (e) at (8, 0) {\Large{R}};
		\node (f) at (10, 2) {\Large{Z}};
		\node (g) at (6, 2) {\Large{V}};
		\node (h) at (0, -1) {\large{\color{brown}Exposure\color{black}}};
		\node (i) at (8, -1) {\large{\color{brown}Incidence\color{black}}};
		\node (j) at (4, -1) {\large{\color{brown}Progression\color{black}}};
		\end{tikzpicture}
	\end{center}
\caption{A plausible causal diagram for a Mendelian randomization study of disease progression.}
\label{Supp.dag2}
\end{figure}

The second reason why we were interested in outcome-selection confounding is because this setting may be relevant for studies of disease progression. It is well known that such studies are affected by selection bias because disease progression phenotypes can only be observed on individuals who have a disease. This is equivalent to conditioning on disease incidence, which induces collider bias between causes of incidence. Figure~\ref{Supp.dag2} illustrates this point for a Mendelian randomization study of disease progression. The outcome of the study is the disease progression trait, while participation is conditional on disease incidence. In such studies it is often implausible to claim that disease progression causes disease incidence. However, there may be unobserved confounders of incidence and progression. Conditional on incidence, such confounders will become associated with the exposure $X$ and result in collider bias.

We performed a simulation to assess whether the pattern of bias is different in the presence of confounding compared to a direct effect of the outcome on the selection coefficient. Our baseline simulation scenario was modified to include the confounder $V$, resulting in the following simulation design:
\begin{eqnarray}
	Z, X, V, \epsilon & \sim & N(0, 1) \nonumber \\
	Y & = & \alpha + \beta  X + \lambda_Y V + \epsilon \nonumber \\
	R & \sim & \text{Logistic} (\pi_R) \nonumber \\
	\text{logit} \left( \pi_R \right) & = & \alpha_R + \beta_R X + \gamma_R Z + \lambda_R V \nonumber
\end{eqnarray}
where $\lambda_Y, \lambda_R$ represent the confounder effects on $Y, R$ respectively. We set $\lambda_Y = \lambda_R = 0.5$ and implemented the simulation for either $\beta = 0.1$ or $\beta = 0$. The other simulation parameters were assigned the same values as in our baseline scenario.

\begin{center}
\begin{table*}[t]%
\caption{Performance of the various selection bias adjustment methods in simulations where the outcome is confounded with selection instead of directly causing it.\label{Supp.table3}}
\centering
\scriptsize
\begin{tabular}{c@{\hskip 0.5in}ccccc@{\hskip 0.5in}cccc}
  \hline
  \multirow{2}{*}{\textbf{Method}} & \multicolumn{5}{c}{\textbf{$\beta = 0.1$}} & \multicolumn{4}{c}{\textbf{$\beta = 0$}} \\
 & \textbf{Mean} & \textbf{Emp SD} & \textbf{StdErr} & \textbf{Cover} & \textbf{Power} & \textbf{Mean} & \textbf{Emp SD} & \textbf{StdErr} & \textbf{Type I} \\ 
  \hline
  CCA 		& 0.075 & 0.016 & 0.016 & 0.666 & 0.996 & -0.025 & 0.016 & 0.016 & 0.341 \\ 
  IPW 		& 0.075 & 0.012 & 0.012 & 0.455 & 1.000 & -0.025 & 0.012 & 0.012 & 0.545 \\ 
  Heckman 	& 0.100 & 0.025 & 0.025 & 0.949 & 0.978 &  0.000 & 0.025 & 0.025 & 0.052 \\ 
  TTW 		& 0.102 & 0.052 & 0.052 & 0.950 & 0.496 &  0.001 & 0.052 & 0.052 & 0.050 \\ 
  Oracle 	& 0.100 & 0.011 & 0.011 & 0.951 & 1.000 &  0.000 & 0.011 & 0.011 & 0.051 \\ 
  \hline
\end{tabular}
\end{table*}
\end{center}

The results are presented in Table~\ref{Supp.table3}. A comparison with our previous results suggests that the effects of selection bias were similar in the two cases. The IVsel methods returned unbiased estimates and confidence intervals with nominal coverage. Bias was still observed for complete-case analysis and IPW, although the bias was somewhat smaller in this case compared to simulations with a direct $Y-R$ effect. This was because the $Y-R$ selection effect was weaker in this simulation, as it was due to confounding and not a direct effect. These results suggest that IVsel methods can be used to account for selection bias in studies of disease progression.

We acknowledge, however, that in practice it can be difficult to identify strong instruments for disease incidence that do not directly affect progression, as would be needed to apply IVsel methods in disease progression studies.

\subsubsection*{Outcome Model Misspecification}

In our final set of simulations for regression analyses, we explored the performance of the selection bias adjustment methods under model misspecification, and in particular when the outcome is continuous but not normally distributed. We did so by altering the distribution of the error term $\epsilon$ in our model specification. Three distributions were considered:
\begin{enumerate}
	\item A $t$ distribution, with four degrees of freedom.
	\item A log-normal distribution with parameters $\mu_{ln} = 0$ and $\sigma^2_{ln} = \frac{1 + \sqrt{5}}{2}$, the latter specified so that the variance of the distribution was equal to $1$.
	\item A mixture of two normal distributions $N(-2, 0.5^2)$ and $N(2, 0.5^2)$ with mixing proportions $0.25$ and $0.75$.
\end{enumerate}
These distributions represent three different forms of misspecification for the outcome. The $t(4)$ distribution is symmetric but has heavier tails than the normal distribution, he log-normal distribution is skewed to the right and the mixture distribution is bimodal. Note that we did not violate the linearity of the outcome model in these simulations, we simply misspecified the error terms.

\begin{center}
\begin{table*}[t]%
\caption{Performance of the various selection bias adjustment methods in simulations with a non-normal outcome and a misspecified outcome model.\label{Supp.table4}}
\centering
\scriptsize
\begin{tabular}{c@{\hskip 0.5in}ccccc@{\hskip 0.5in}cccc}
  \hline
  \textbf{Method} & \textbf{Mean} & \textbf{Emp SD} & \textbf{StdErr} & \textbf{Cover} & \textbf{Power} & \textbf{Mean} & \textbf{Emp SD} & \textbf{StdErr} & \textbf{Type I} \\ 
  \hline
   & \multicolumn{9}{c}{$t(4)$ distribution} \\
  \hline
  CCA 		&  0.022 & 0.020 & 0.020 & 0.030 & 0.208 & -0.072 & 0.020 & 0.020 & 0.950 \\ 
  IPW 		&  0.021 & 0.016 & 0.016 & 0.000 & 0.264 & -0.074 & 0.016 & 0.016 & 0.998 \\ 
  Heckman 	&  0.100 & 0.033 & 0.033 & 0.956 & 0.871 &  0.000 & 0.031 & 0.031 & 0.047 \\ 
  TTW 		&  0.114 & 0.064 & 0.063 & 0.941 & 0.444 &  0.013 & 0.065 & 0.064 & 0.055 \\ 
  Oracle 	&  0.100 & 0.014 & 0.014 & 0.950 & 1.000 &  0.000 & 0.014 & 0.014 & 0.052 \\ 
  \hline
   & \multicolumn{9}{c}{Log-normal distribution} \\
  \hline  
  CCA 		&  0.033 & 0.017 & 0.017 & 0.028 & 0.489 & -0.061 & 0.017 & 0.017 & 0.940 \\ 
  IPW 		&  0.049 & 0.011 & 0.012 & 0.004 & 0.980 & -0.046 & 0.011 & 0.011 & 0.988 \\ 
  Heckman 	&  0.102 & 0.029 & 0.029 & 0.950 & 0.946 &  0.001 & 0.027 & 0.027 & 0.052 \\ 
  TTW 		&  0.132 & 0.053 & 0.053 & 0.907 & 0.697 &  0.030 & 0.056 & 0.055 & 0.086 \\ 
  Oracle 	&  0.100 & 0.010 & 0.010 & 0.951 & 1.000 &  0.000 & 0.010 & 0.010 & 0.047 \\ 
  \hline
   & \multicolumn{9}{c}{Bimodal distribution} \\
  \hline
  CCA 		&  0.022 & 0.022 & 0.022 & 0.054 & 0.170 & -0.074 & 0.022 & 0.022 & 0.918 \\ 
  IPW 		& -0.003 & 0.019 & 0.020 & 0.001 & 0.047 & -0.098 & 0.019 & 0.019 & 0.999 \\ 
  Heckman 	&  0.097 & 0.036 & 0.036 & 0.950 & 0.781 & -0.002 & 0.035 & 0.034 & 0.050 \\ 
  TTW 		&  0.075 & 0.072 & 0.069 & 0.923 & 0.203 & -0.022 & 0.073 & 0.071 & 0.067 \\ 
  Oracle 	&  0.100 & 0.018 & 0.018 & 0.949 & 1.000 &  0.000 & 0.018 & 0.018 & 0.051 \\ 
  \hline
\end{tabular}
\end{table*}
\end{center}

We implemented the various selection bias adjustment methods with a normally distributed outcome model. The results are reported in Table~\ref{Supp.table4} and can be compared with the baseline simulation in Table 1 in the main part of the paper. Heckman's method was the least affected by misspecifying the error term in this simulation, and produced practically unbiased parameter estimates. The TTW method was subject to a small degree of bias, while for the CCA and IPW methods, the bias was slightly larger than in our baseline simulation. These findings were observed consistently across the three distributional assumptions for the outcome.

\subsection*{Mendelian Randomization}

We now report the results of additional simulations conducted in the Mendelian randomization setting, to complement the results reported in the paper. In particular, we consider four sets of simulations. First, we report results for one-sample MR simulations with multiple instruments for inference, obtained by computing the selection-adjusted summary statistics. Second, we focus on one-sample MR with a single instrument for inference and vary other aspects of the MR simulation design to explore how they affect the IVsel methods. Third, we explore the power of IVsel methods to detect a causal effect. And fourth, we investigate how the various selection bias adjustment methods are affected by misspecification of the first-stage regression model.

\subsubsection*{Summary Statistics for One-sample MR}

In the main part of our paper, we discussed two approaches for one-sample MR with multiple instruments for inference. The first approach was based on a modification of the 2SLS method. A second approach was to use the IVsel methods for each genetic instrument for inference separately, compute summary statistics adjusted for selection bias and combine them using the IVW formula. This approach was used in our two-sample MR simulations in the main part of our paper, but there is nothing preventing us from utilizing it in a one-sample setting apart from the fact that using the IVW formula in one-sample MR can induce bias due to sample overlap. This bias is only likely to appear in weak instrument settings though, and at the same time the summary statistics approach is computationally easier to implement. Hence, it may be preferable in simulations with strong instruments where bias due to sample overlap is less likely.

We decided to explore the performance of the summary statistics approach in one-sample MR simulations. For that purpose, we used the same data as those to which we fitted the 2SLS method (see Table~4 of the main part of the paper). For comparison, we implemented a summary-statistics approach not only for the IVsel methods but also for complete-case analysis and IPW.  

\begin{center}
\begin{table*}[t]%
\caption{Performance of the various selection bias adjustment methods in one-sample Mendelian randomization simulations with multiple genetic instruments for inference, using a summary statistics approach instead of 2SLS.\label{Supp.table5}}
\centering
\scriptsize
\begin{tabular}{c@{\hskip 0.5in}ccccc@{\hskip 0.5in}cccc}
  \hline
  \multirow{2}{*}{\textbf{Method}} & \multicolumn{5}{c}{\textbf{$\theta = 0.2$}} & \multicolumn{4}{c}{\textbf{$\theta = 0$}} \\
   & \textbf{Causal} & \textbf{Emp SD} & \textbf{StdErr} & \textbf{Cover} & \textbf{Power} & \textbf{Causal} & \textbf{Emp SD} & \textbf{StdErr} & \textbf{Type I} \\ 
  \hline
   & \multicolumn{9}{c}{One-sample MR (Summary Statistics) \;\; - \;\; $Y \rightarrow R$} \\
  \hline
  CCA 		& 0.147 & 0.054 & 0.061 & 0.900 & 0.686 &  0.008 & 0.051 & 0.057 & 0.025 \\ 
  IPW 		& 0.147 & 0.038 & 0.049 & 0.889 & 0.911 &  0.006 & 0.037 & 0.044 & 0.026 \\ 
  Heckman 	& 0.207 & 0.059 & 0.067 & 0.975 & 0.901 &  0.010 & 0.055 & 0.061 & 0.030 \\ 
  TTW 		& 0.231 & 0.338 & 0.368 & 0.968 & 0.076 &  0.004 & 0.299 & 0.320 & 0.040 \\ 
  Oracle 	& 0.206 & 0.044 & 0.051 & 0.974 & 0.990 &  0.008 & 0.043 & 0.046 & 0.046 \\ 
  \hline
   & \multicolumn{9}{c}{One-sample MR (Summary Statistics) \;\; - \;\; $X \rightarrow R$} \\
  \hline
  CCA 		& 0.273 & 0.058 & 0.068 & 0.860 & 0.994 &  0.009 & 0.057 & 0.061 & 0.035 \\ 
  IPW 		& 0.276 & 0.056 & 0.067 & 0.845 & 0.996 &  0.009 & 0.057 & 0.061 & 0.036 \\ 
  Heckman 	& 0.206 & 0.043 & 0.051 & 0.980 & 0.994 &  0.007 & 0.043 & 0.046 & 0.035 \\ 
  TTW 		& 0.113 & 0.045 & 0.045 & 0.492 & 0.719 &  0.005 & 0.032 & 0.035 & 0.039 \\ 
  Oracle 	& 0.208 & 0.042 & 0.050 & 0.978 & 0.996 &  0.007 & 0.043 & 0.046 & 0.036 \\ 
  \hline
   & \multicolumn{9}{c}{One-sample MR (Summary Statistics) \;\; - \;\; $X, Y \rightarrow R$} \\
  \hline
  CCA 		& 0.077 & 0.067 & 0.073 & 0.625 & 0.157 & -0.103 & 0.065 & 0.067 & 0.319 \\ 
  IPW 		& 0.067 & 0.048 & 0.057 & 0.344 & 0.163 & -0.109 & 0.045 & 0.051 & 0.570 \\ 
  Heckman 	& 0.214 & 0.062 & 0.068 & 0.962 & 0.902 &  0.010 & 0.061 & 0.062 & 0.046 \\ 
  TTW 		& 0.382 & 0.229 & 0.285 & 0.953 & 0.225 &  0.192 & 0.227 & 0.255 & 0.091 \\ 
  Oracle 	& 0.208 & 0.043 & 0.051 & 0.976 & 0.991 &  0.007 & 0.042 & 0.046 & 0.034 \\ 
  \hline
\end{tabular}
\end{table*}
\end{center}

Table~\ref{Supp.table5} contains the results of this simulation experiment. A comparison with the results obtained from the 2SLS method suggests that the impact of sample overlap was rather small. This is not unexpected, since the effect of sample overlap usually manifests as an increase in weak instrument bias. Our MR simulations were parametrized to avoid weak instrument bias and the corresponding $F$ statistics were much larger than the threshold of $10$ that is commonly used to detect weak instruments, therefore the impact of weak instrument bias was small. We also note that the performance of the TTW method did not deteriorate further compared to the 2SLS implementation of the method. The only notable difference between 2SLS and the summary statistics approach occurred for the complete-case analysis and IPW methods when selection was a function of the exposure. In this case, selection bias acted in opposite directions for the summary statistics approach compared to the 2SLS method, as we discussed in the main part of our manuscript.

\subsubsection*{Additional MR Simulations}

In the case of regression analyses, we have covered a wide range of simulation scenarios, varying the values of several simulation parameters. Some conclusions about the performance of IVsel methods in MR settings can be extracted from these regression-based simulations, as we have already mentioned. Here, we explored some of these scenarios explicitly in an MR setting. Specifically, we performed four new sets of simulations. First, we varied the proportion of individuals in the sample with missing data. Second, we conducted simulations with different sample sizes to assess whether the IVsel methods can be used in small-sample MR analyses. And third, we considered violations of the instrumental variable assumptions for selection by inducing direct effects of the instrument for selection on the MR exposure or outcome.

To reduce the number of simulations conducted, we focused on one-sample MR analyses with a single instrument for inference (which could be a polygenic risk score). We also assumed that both the exposure and the outcome affect selection into the MR study, and that both are missing for individuals with missing data. Therefore, the simulations included in this section can be seen as extensions of (and thus compared to) the third set of results in Table~3 of the main part of the manuscript.

The simulations of this section were generated as follows. For the proportion of missingness, we varied the parameter $\alpha_R$ in the selection model, letting it take values that correspond to $10\%$, $20\%$, $35\%$, $50\%$, $65\%$, $80\%$, $90\%$ of individuals having fully observed data. For simulations with different sample sizes, we used the following values: $N = 10^3$, $2 \times 10^3$, $5 \times 10^3$, $10^4$, $2 \times 10^4$, $5 \times 10^4$, $10^5$; simulations with different sample sizes were only repeated $10^3$ times for each $N$ value to reduce computational cost. For simulations with a $Z-X$ effect, we modified the generation of exposure data to $X = \alpha_X + \beta_X G + \gamma_X U + \lambda_X Z + \epsilon_X$, using the values $\lambda_X = 0, 0.1, 0.2, 0.3, 0.4, 0.5$ for the instrument-exposure parameter. For a $Z-Y$ effect, we generated outcome data from $Y = \alpha_Y + \theta X + \gamma_Y U + \lambda_Y Z + \epsilon_Y$, using $\lambda_Y = 0, 0.1, 0.2, 0.3, 0.4, 0.5$ as possible values.

\begin{figure}[h]
\begin{center}
	\includegraphics[scale = 0.67]{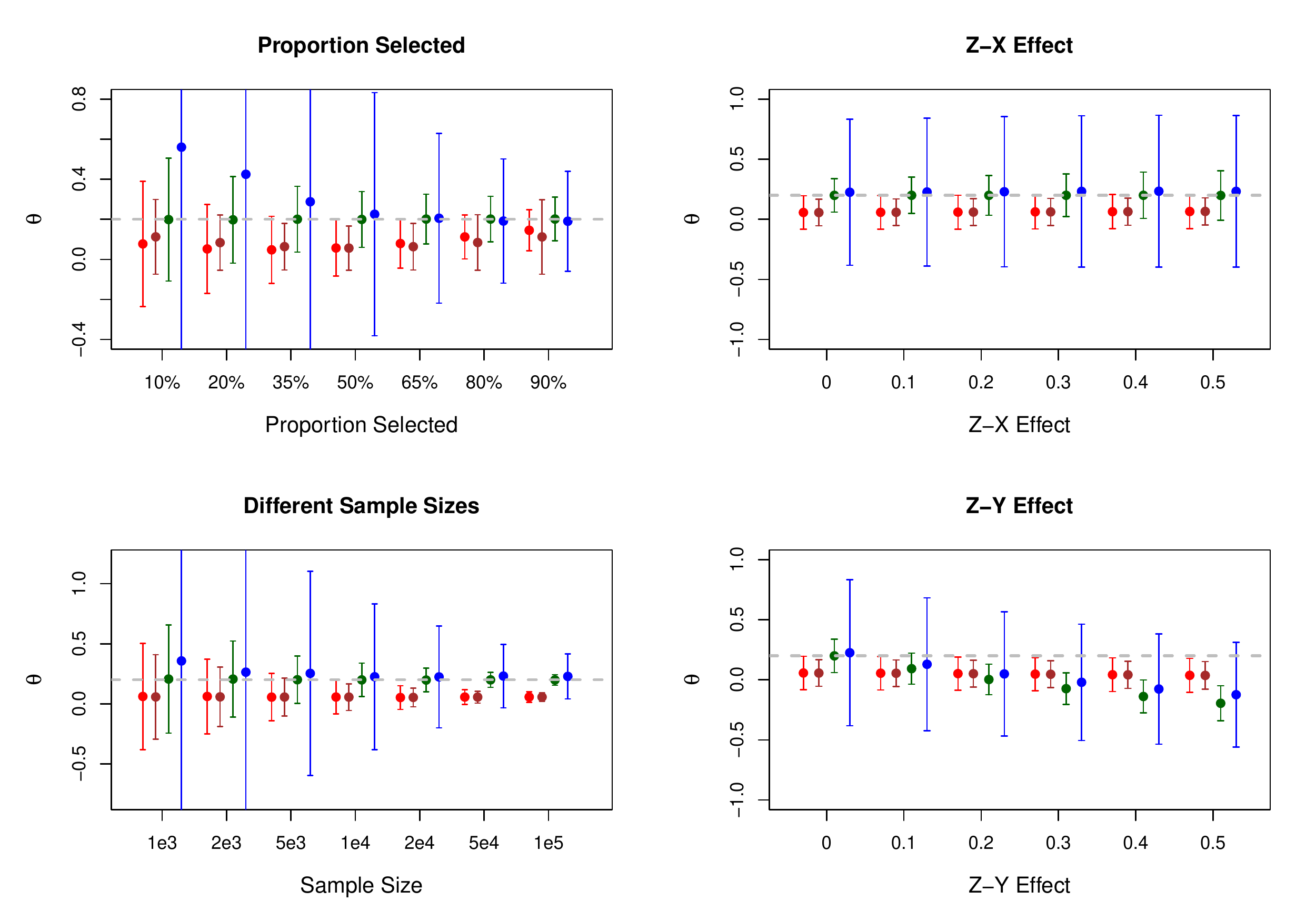}
\end{center}
\caption{Performance of the various selection bias adjustment methods in MR analyses with a varying proportion of missing data (top left), an added $Z-X$ effect (top right), different sample sizes (bottom left), or an added $Z-Y$ effect (bottom right).}
\label{Supp.figure1.2}
\end{figure}

The results of these additional simulations are summarized in Figure~\ref{Supp.figure1.2}. To improve the presentation of our results, we have reported median (and not mean) causal effect estimates and computed the averaged confidence intervals using the median (and not mean) standard errors for each method.

Varying the proportion of individuals with fully observed data had a similar impact in MR simulations as in regression-based ones: smaller proportions meant that inference was based on fewer individuals, hence the precision of all methods was reduced. This was more prominent for the TTW method, whose precision was already quite low in the baseline simulation where $50\%$ of all individuals had fully observed data. With a high proportion of missing values, MR analyses would have to be conducted in a sample of only a few thousand individuals, and the use of a method with high in-built uncertainty, such as TTW, seems infeasible in these settings. On the other hand, confidence intervals produced by Heckman's method had comparable width to those from complete-case analysis or IPW.

Similar results were obtained in simulations with different sample sizes. When the sample size of the MR analysis was small, e.g. a few thousand individuals) the various methods had substantial uncertainty in their causal effect estimates. Again, this was more problematic for the TTW method due to its lower overall precision. The complete-case, IPW and Heckman methods had comparable standard errors. On the other hand, with larger sample sizes, the performance of all methods improved.

Unlike the regression-based simulations, a $Z-X$ effect does constitute a violation of the exclusion restriction for $Z$ because the MR exposure acts as an ``outcome" for the purpose of estimating the denominator of the Wald ratio. However, inducing a direct $Z-X$ effect did not seem to have a substantial impact on IVsel estimates in our simulations. Note that a $Z-X$ effect violates the exclusion restriction in both the $G-X$ and the $G-Y$ regression analyses (since $Z \notindep Y | G$), inducing bias in both. Our results in Figure~\ref{Supp.figure1.2} suggest that these two biases may cancel out when computing the Wald ratio estimate. 

Finally, with a direct effect between the instrument $Z$ and the outcome $Y$, the second IV assumption for $Z$ was violated and bias was induced. The pattern of bias was similar to that observed in Figure~\ref{Supp.figure1.1} for regression-based analyses, though its magnitude was somewhat weaker. Again, this means that $Z-Y$ effects can devalidate the results of the IVsel methods and this should be kept in mind when deciding which variables to use as instruments for selection. Thankfully, it may be possible to detect such effects in some applications. For example, consider an MR analysis where the instrument for selection is genetic. With access to data from an outcome GWAS, the researchers can investigate whether the instrument for selection $Z$ correlates with the outcome conditional on any instruments for inference $G$, by looking at the associations between the genetic variants that constitute $Z$ and the outcome in the GWAS. Note that the outcome GWAS will itself be affected by the missing data, so this idea comes with no theoretical guarantees of validity, but it could be an informal heuristic for detecting variants in $Z$ that are strongly associated with the outcome and removing them from the instrument for selection.

\subsubsection*{Empirical Power of IVsel Methods}

In addition to the previous scenarios, we also conducted an MR simulation varying the value of the exposure-outcome causal effect $\theta$ in order to study the power of the IVsel methods to detect a non-null causal effect. Specifically, we considered the values $\theta = 0$, $0.05$, $0.1$, $0.15$, $0.2$, $0.25$, $0.3$ for the causal effect. Once again, this was done only for the case of a one-sample MR analysis with a single instrument for inference, where both the exposure and the outcome have missing data.

\begin{figure}[h]
\begin{center}
	\includegraphics[scale = 0.67]{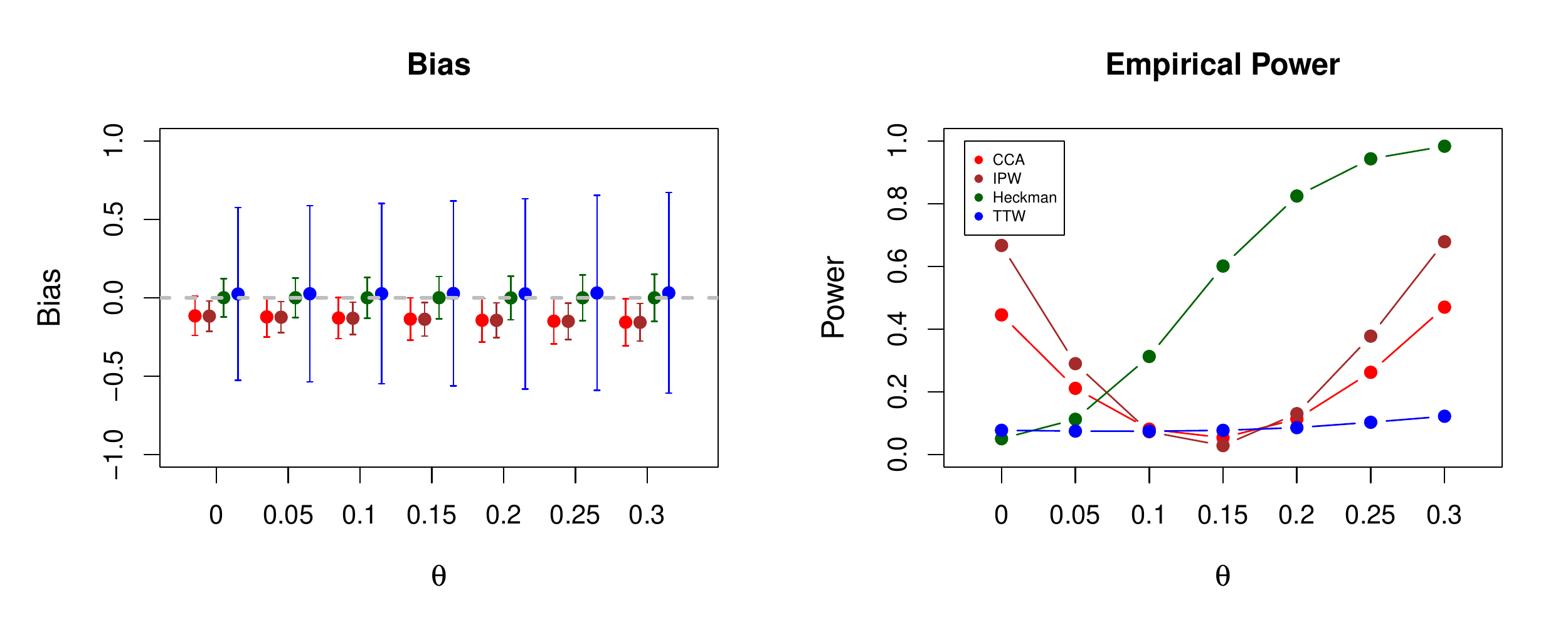}
\end{center}
\caption{Performance of the various selection bias adjustment methods in MR analyses with different values of the causal effect $\theta$, in terms of causal effect estimation (left) and empirical power to reject the causal null hypothesis (right).}
\label{Supp.figure1.3}
\end{figure}

Figure~\ref{Supp.figure1.3} illustrates the results of this simulation. We plot median causal effect estimates (left) and the empirical power of each method to reject the causal null hypothesis (right). The results suggest that varying the value of the causal effect parameter had little impact on the performance of the various methods. However, the difference in the power of the two IVsel methods was considerable. Heckman's method had $60.1\%$ empirical power to detect an effect of $0.15$, $82.5\%$ power to detect an effect of $0.2$ and $94.3\%$ power to detect an effect of $0.25$. On the other hand, the TTW method struggled for power and only had $10.3\%$ power to detect an effect of $0.25$. It is worth mentioning that these results concern an MR analysis with a sample size of $N = 10000$ (with approximately half of those individuals having missing data), therefore low power is not unexpected. Nevertheless, the power of TTW is considerably lower than that of the Heckman method, and this should be taken into account when deciding whether to use it in applications.

\subsubsection*{MR with a Misspecified First-stage Regression Model}

Finally we discuss the impact of model misspecification in the first-stage regression on the performance of selection bias adjustment methods. In particular, we considered the setting where a linear model is assumed for the $G-X$ association, while the true association is non-linear. \cite{Zhao2019} reported that such misspecification does not introduce bias in a one-sample MR analysis or in a two-sample analysis where the two samples come from the same population. However, in a two-sample analysis where the two samples come from different populations, the misspecification can be a cause of concern.

Here, we aimed to extend the literature results and assess whether the IVsel methods assuming a linear $G-X$ association are indeed robust to model misspecification in the first-stage regression. With that in mind, we implemented six simulation scenarios:
\begin{enumerate}
	\item One-sample MR, linear first stage.
	\item One-sample MR, non-linear first stage.
	\item Two-sample MR, same population, linear first stage.
	\item Two-sample MR, same population, non-linear first stage.
	\item Two-sample MR, different populations, linear first stage.
	\item Two-sample MR, different populations, non-linear first stage.
\end{enumerate}
For simplicity, we used a single instrument for inference in this simulation and assumed that missingness was only affected by the outcome $Y$. In simulations with a linear first stage, exposure values were generated according to the equation
\begin{equation*}
	X = \beta_X G + \gamma_X U + \epsilon_{X}
\end{equation*}
In scenarios with a non-linear first stage, exposure values were generated according to the equation
\begin{equation*}
	X = \beta_X G + \beta_{X_2} G^2 + \gamma_X U + \epsilon_{X}
\end{equation*}
adding a quadratic effect of the instrument $G$ on exposure. Values for $G$ were generated from a $N(0, 1)$ distribution by default. In the last two scenarios where there were two distinct populations, we generated genetic instrument measurements from a $N(0,1)$ distribution for the first sample and from a $t(4)$ distribution for the second sample. The value of the additional parameter $\beta_{X_2}$ was set to $0.5$, representing a fairly strong non-linear effect, to better demonstrate the impact of model misspecification. The true causal effect was set to $\theta = 0.2$.

\begin{center}
\begin{table*}[t]%
\caption{Performance of the various selection bias adjustment methods in Mendelian randomization simulations with a misspecified first-stage model.\label{nl.sim1}}
\centering
\scriptsize
\begin{tabular}{c@{\hskip 0.5in}ccccc@{\hskip 0.5in}ccccc}
  \hline
  \multirow{2}{*}{\textbf{Method}} & \multicolumn{5}{c}{\textbf{Linear 1st Stage}} & \multicolumn{5}{c}{\textbf{Non-linear 1st Stage}} \\
   & \textbf{Causal} & \textbf{Emp SD} & \textbf{StdErr} & \textbf{Cover} & \textbf{Power} & \textbf{Causal} & \textbf{Emp SD} & \textbf{StdErr} & \textbf{Cover} & \textbf{Power} \\ 
  \hline
   & \multicolumn{10}{c}{One-sample MR} \\
  \hline
  CCA 		& 0.143 & 0.056 & 0.059 & 0.847 & 0.696 &  0.154 & 0.054 & 0.057 & 0.882 & 0.787 \\ 
  IPW 		& 0.143 & 0.039 & 0.049 & 0.847 & 0.886 &  0.142 & 0.038 & 0.050 & 0.860 & 0.875 \\ 
  Heckman 	& 0.199 & 0.061 & 0.065 & 0.962 & 0.882 &  0.209 & 0.060 & 0.063 & 0.958 & 0.920 \\ 
  TTW 		& 0.203 & 0.343 & 0.340 & 0.947 & 0.094 &  0.545 & 0.361 & 0.319 & 0.782 & 0.416 \\ 
  Oracle 	& 0.199 & 0.044 & 0.049 & 0.971 & 0.986 &  0.199 & 0.044 & 0.050 & 0.972 & 0.987 \\ 
  \hline
   & \multicolumn{10}{c}{Two-sample MR - Same Population} \\
  \hline
  CCA 		& 0.144 & 0.058 & 0.059 & 0.841 & 0.690 &  0.155 & 0.058 & 0.057 & 0.873 & 0.774 \\ 
  IPW 		& 0.144 & 0.041 & 0.049 & 0.835 & 0.875 &  0.144 & 0.043 & 0.050 & 0.835 & 0.853 \\ 
  Heckman 	& 0.200 & 0.064 & 0.065 & 0.950 & 0.873 &  0.211 & 0.067 & 0.063 & 0.932 & 0.906 \\ 
  TTW 		& 0.213 & 0.344 & 0.340 & 0.946 & 0.101 &  0.551 & 0.368 & 0.319 & 0.773 & 0.415 \\ 
  Oracle 	& 0.200 & 0.049 & 0.049 & 0.951 & 0.984 &  0.201 & 0.052 & 0.050 & 0.941 & 0.980 \\ 
  \hline
   & \multicolumn{10}{c}{Two-sample MR - Different Populations} \\
  \hline
  CCA 		& 0.144 & 0.042 & 0.042 & 0.738 & 0.936 &  0.191 & 0.834 & 0.049 & 0.365 & 0.780 \\ 
  IPW 		& 0.144 & 0.030 & 0.035 & 0.676 & 0.995 &  0.024 & 0.762 & 0.216 & 0.895 & 0.083 \\ 
  Heckman 	& 0.201 & 0.047 & 0.047 & 0.949 & 0.992 &  0.241 & 0.843 & 0.054 & 0.390 & 0.814 \\ 
  TTW 		& 0.214 & 0.214 & 0.211 & 0.946 & 0.175 &  4.168 & 2.955 & 0.302 & 0.022 & 0.985 \\ 
  Oracle 	& 0.200 & 0.035 & 0.036 & 0.952 & 1.000 &  0.213 & 0.671 & 0.043 & 0.435 & 0.861 \\ 
  \hline
\end{tabular}
\end{table*}
\end{center}

Results are summarized in Table~\ref{nl.sim1}. For simulations with a linear first-stage model and no population differences, the performance of the various methods has already been reported in Table 3 of the main part of the paper and is repeated here. The two-sample MR simulation with different populations and a linear first-stage regression suggests that population differences only had a small impact on this set of simulations. The CCA and IPW methods were still biased, but their bias was similar to that in simulations without population differences, suggesting that it is due to missing data. An increase in the precision of causal effect estimates was observed for all methods. 

Our goal here was to compare the performance of the various methods between simulations with a linear and a non-linear first-stage model. For one-sample MR, this comparison suggested that the TTW method was quite sensitive to misspecification of the first-stage regression. Heckman's algorithm and complete-case analysis were affected to a lesser extent (for CCA the overall bias was smaller for a misspecified first-stage model because bias due to misspecification and bias due to missing data acted in opposite directions here). Oracle estimates were unaffected by the misspecification, as suggested in the literature \citep{Zhao2019}, and so were the IPW estimates.

For two-sample MR with data from the same population, the effect of non-linearity in the first-stage regression was similar to that in one-sample MR. However, for two-sample MR with data from different populations, the effect was more pronounced. TTW was again the most impacted method, with causal effect estimates taking rather extreme values in some iterations. Heckman's method and CCA were also substantially impacted, while oracle estimates were biased but to a smaller degree. Interestingly, the combination of population differences and misspecification in the weighting model had a strong effect on the performance of IPW. The empirical standard deviation of estimates was larger for all methods, while their coverage was also compromised.

The scope of the current simulation was quite limited, as we only considered one form of first-stage model misspecification (a quadratic term for $G$ that is present in the data-generating model but is omitted from the selection model) and one way of generating data from different populations (genetic data simulated from normal and $t(4)$ distributions for the two populations respectively). Therefore we cannot claim our results to be representative of model misspecification and population differences in general and more research in that direction is needed. However, our simulation suggests that first-stage misspecification and population differences are more concerning when acting simultaneously: when only one of these issues was present, the bias of most methods was fairly small in absolute value. An exception occurred for the TTW method, which was more sensitive to misspecification of the first-stage regression than other methods.

\section*{Additional Results from the Real-data Application}

Here, we report additional diagnostics and results from our real-data application. In the main part of our manuscript, we focused on the effects of BMI on smoking status and smoking frequency. Here, we expand this analysis in a number of ways. First, we provide more details on the construction of the polygenic risk score for ALSPAC participation. Second, we report numerical results for the two smoking outcomes, to complement the plot in Figure 3 of the main part, as well as a sensitivity analysis for the number of cigarettes smoked, conducted using Poisson regression. Third, we consider the effects of BMI on two additional behavioral outcomes whose measurement is often subject to selection bias, namely self-harm and depression status. And fourth, we report measures of instrument strength for each instrument for selection. 

\subsection*{Construction of the Risk Score for ALSPAC Participation}

Here, we provide more details on the construction of the polygenic risk score for ALSPAC participation. The construction was based on previous work by \cite{Taylor2018}, who investigated genetic associations with participation in ALSPAC. We used data from their work, kindly provided to us by the authors of the study. We focused on overall children participation in order to construct the score, since our MR analyses use data from both clinic visits (BMI) and questionnaires (smoking traits).

Only nine genetic variants were identified as having a GWAS-significant association with overall children participation \citep{Taylor2018}. These variants were all in high LD ($> 99\%$) and were located in the gene region \textit{BDKRB1} in Chromosome 14. The leading variant was rs28631073. Using only that one variant would produce a rather weak instrument for selection. Therefore, we decided to create a more inclusive polygenic score for ALSPAC participation by relaxing the threshold for inclusion of genetic variants into the risk score. Specifically, we included all variants whose p-value of association with ALSPAC participation was lower than $10^{-4}$. This yielded a total of $1272$ variants. The number reduced to $1100$ after quality control (discarding non-replacement alleles, triallelic SNPs and variants with minor allele frequency below $1\%$). We then implemented LD-pruning at a correlation threshold of $0.1$ to remove correlated variants; this reduced the number of variants to $178$. 

We assessed potential violations of the IV assumptions by examining the associations of variants in our polygenic score with BMI, smoking status and the number of cigarettes smoked per day. The associations were quite weak, and in fact none of the $178$ variants was associated with any of the three traits after applying a correction for multiple testing. Therefore, all $178$ variants were suitable for inclusion into the risk score.

Our PRS was constructed using the associations of the $178$ variants with ALSPAC participation as weights. We acknowledge the potential for winner's curse bias to have affected the construction of our risk score, unlike the scores constructed using variants from the UK Biobank. However, this is a price one may have to pay when using genetic instruments for selection. Factors that affect participation in ALSPAC are slightly different to factors affecting participation in UK Biobank or other datasets, so identifying variants for ALSPAC participation based on patterns of participation in other datasets will also be subject to biases. Accordingly, the risk score for participation constructed using ALSPAC data was a much stronger instrument for selection compared to the two risk scores based on the UK Biobank, as we discuss in the next subsections.

\subsection*{Effects of BMI on Smoking Traits Revisited}

Here, we provide numerical results for the effects of BMI on smoking traits; these are reported in Table~\ref{alspac.table1}. Figure 3 in the main part of the paper was created based on these results. As we have already discussed, results obtained from the IVsel methods agree with those from complete-case analysis and IPW in suggesting a causal effect of BMI on number of cigarettes smoked per day, but no effect on smoking status.

\begin{center}
\begin{table*}[t]%
\caption{MR estimates of the effects of BMI on smoking status and number of cigarettes per day in ALSPAC, estimated using various methods and instruments for selection to adjust for missing data.\label{alspac.table1}}
\centering
\scriptsize
\begin{tabular}{cc@{\hskip 0.5in}ccc@{\hskip 0.5in}ccc}
  \hline
   \multirow{2}{*}{\textbf{Method}} & \multirow{2}{*}{\textbf{Instrument}} & \multicolumn{3}{c}{\textbf{Smoking Status}} & \multicolumn{3}{c}{\textbf{Number of Cigarettes}} \\
   & & \textbf{Causal} & \textbf{SD} & \textbf{$95\%$ CI} & \textbf{Causal} & \textbf{SD} & \textbf{$95\%$ CI} \\ 
  \hline
  CCA 		& ----- 	&  0.028 & 0.048 & (-0.065 , 0.122) & 0.226 & 0.087 & ( 0.056 , 0.396) \\ 
  IPW 		& ----- 	&  0.044 & 0.029 & (-0.013 , 0.102) & 0.182 & 0.089 & ( 0.007 , 0.356) \\ 
  Heckman 	& RCT 		&  0.005 & 0.030 & (-0.053 , 0.063) & 0.215 & 0.097 & ( 0.026 , 0.405) \\ 
  Heckman 	& ALSPAC 	&  0.017 & 0.030 & (-0.041 , 0.075) & 0.227 & 0.087 & ( 0.057 , 0.398) \\ 
  Heckman 	& FFQ 		&  0.016 & 0.027 & (-0.037 , 0.069) & 0.206 & 0.088 & ( 0.033 , 0.380) \\ 
  Heckman 	& MHQ 		&  0.015 & 0.026 & (-0.037 , 0.067) & 0.204 & 0.087 & ( 0.033 , 0.375) \\ 
  TTW 		& RCT 		& -0.039 & 0.144 & (-0.320 , 0.243) & 0.666 & 0.828 & (-0.958 , 2.289) \\ 
  TTW 		& ALSPAC 	& -0.043 & 0.118 & (-0.275 , 0.189) & 0.541 & 0.484 & (-0.407 , 1.488) \\ 
  TTW 		& FFQ 		&  0.008 & 0.144 & (-0.274 , 0.290) & 1.027 & 1.408 & (-1.732 , 3.786) \\ 
  TTW 		& MHQ 		&  0.017 & 0.140 & (-0.257 , 0.291) & 0.830 & 1.143 & (-1.410 , 3.070) \\ 
  \hline
\end{tabular}
\end{table*}
\end{center}

\begin{figure}[ht]
\begin{center}
	\includegraphics[scale = 0.52]{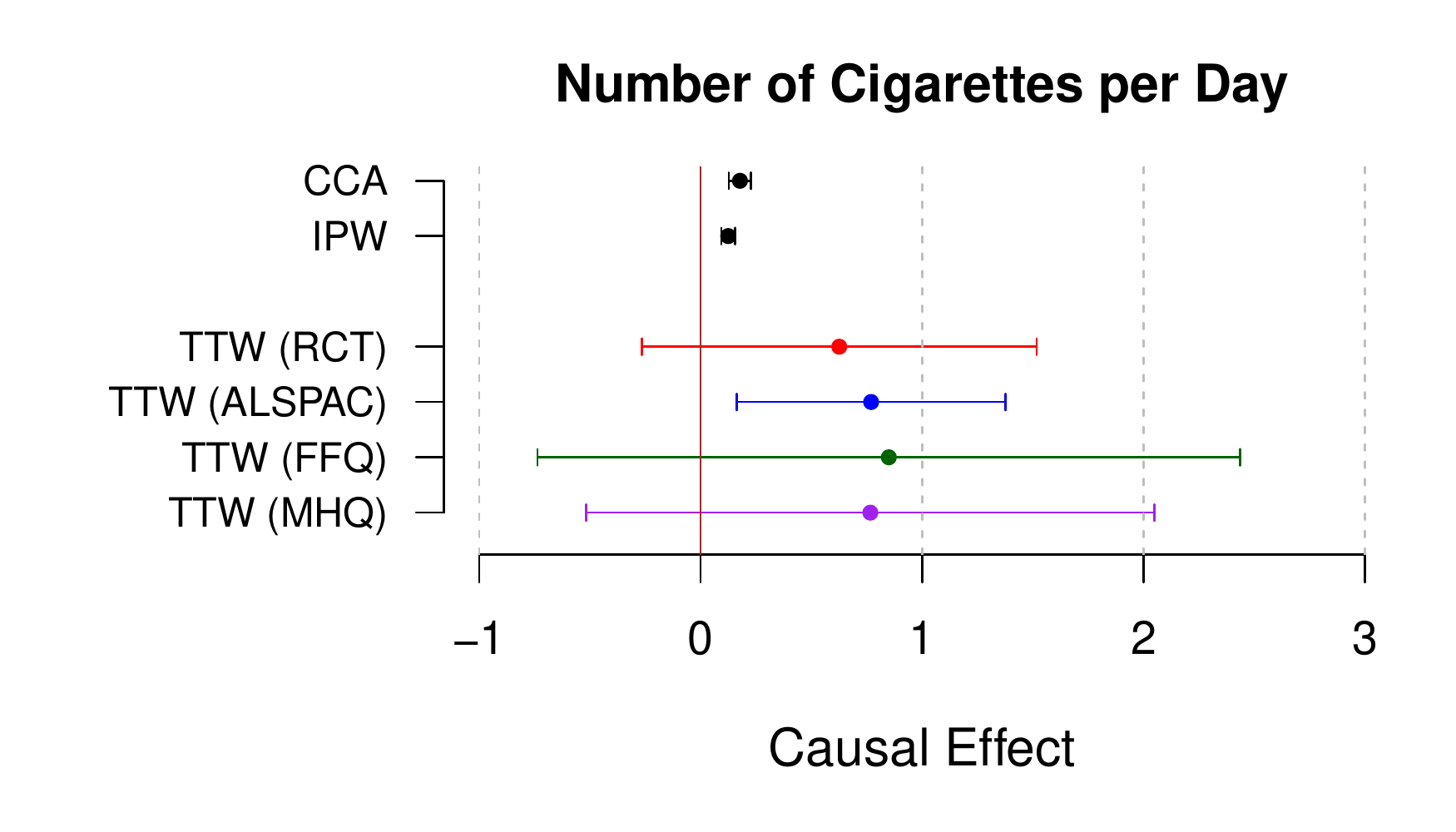}
\end{center}
\caption{MR estimates of the effect of BMI on the number of cigarettes smoked per day, obtained from ALSPAC data using Poisson regression to model the number of cigarettes smoked and various methods to adjust for missing data. Colors indicate which instrument for selection is used each time (red: RCT for participation, blue: ALSPAC risk score, green: SNPs associated with FFQ completion in UK Biobank, purple: SNPs associated with MHQ completion in UK Biobank, black: method requires no instrument for selection).}
\label{alspac.supp.fig1}
\end{figure}

The MR analysis for the number of cigarettes smoked per day was conducted using linear regression, in order to demonstrate the use of the various selection bias adjustment methods. However, it is also possible (and indeed more reasonable) to model the number of cigarettes smoked per day using Poisson regression. Hence, we modified the MR analyses reported in the main part of the paper by estimating SNP-outcome associations from Poisson regression instead of linear regression, and report Wald ratios, model-based standard errors and $95\%$ confidence intervals for each method in Table~\ref{alspac.table2}. Results are not reported for Heckman's sample selection method, because the implementation of the method in \texttt{R} does not support a Poisson model. A graphical illustration of the results is provided in Figure~\ref{alspac.supp.fig1}. Briefly, the results obtained using the Poisson regression model agree with those from linear regression in advocating a risk-increasing effect of BMI on smoking frequency, although again the TTW method is underpowered and produces wide confidence intervals.

\begin{center}
\begin{table*}[t]%
\caption{MR estimates of the effects of BMI on number of cigarettes smoked in ALSPAC, estimated using Poisson regression, and various methods and instruments for selection to adjust for missing data. Estimates from Heckman's method are not reported, because the existing software does not support an implementation of the method for Poisson regression.\label{alspac.table2}}
\centering
\scriptsize
\begin{tabular}{cc@{\hskip 0.5in}ccc}
  \hline
   \multirow{2}{*}{\textbf{Method}} & \multirow{2}{*}{\textbf{Instrument}} & \multicolumn{3}{c}{\textbf{Number of Cigarettes (Poisson)}} \\
   & & \textbf{Causal} & \textbf{SD} & \textbf{$95\%$ CI} \\ 
  \hline
  CCA 		& ----- 	& 0.178 & 0.025 & ( 0.128 , 0.227) \\
  IPW 		& ----- 	& 0.124 & 0.016 & ( 0.094 , 0.155) \\
  TTW 		& RCT 		& 0.626 & 0.455 & (-0.265 , 1.517) \\
  TTW 		& ALSPAC 	& 0.770 & 0.309 & ( 0.163 , 1.376) \\ 
  TTW 		& FFQ 		& 0.850 & 0.809 & (-0.736 , 2.435) \\
  TTW 		& MHQ 		& 0.766 & 0.654 & (-0.516 , 2.048) \\ 
  \hline
\end{tabular}
\end{table*}
\end{center}

\subsection*{Effects of BMI on Self-Harm and Depression Status}

In addition to our analysis about the effects of BMI on smoking traits, we also considered two additional behavioral outcomes, these being self-harm and depression status. Both traits are likely to be affected by selection bias due to missing data, since study participants are often reluctant to disclose personal information about their mental health.

\begin{figure}[ht]
\begin{center}
	\includegraphics[scale = 0.45]{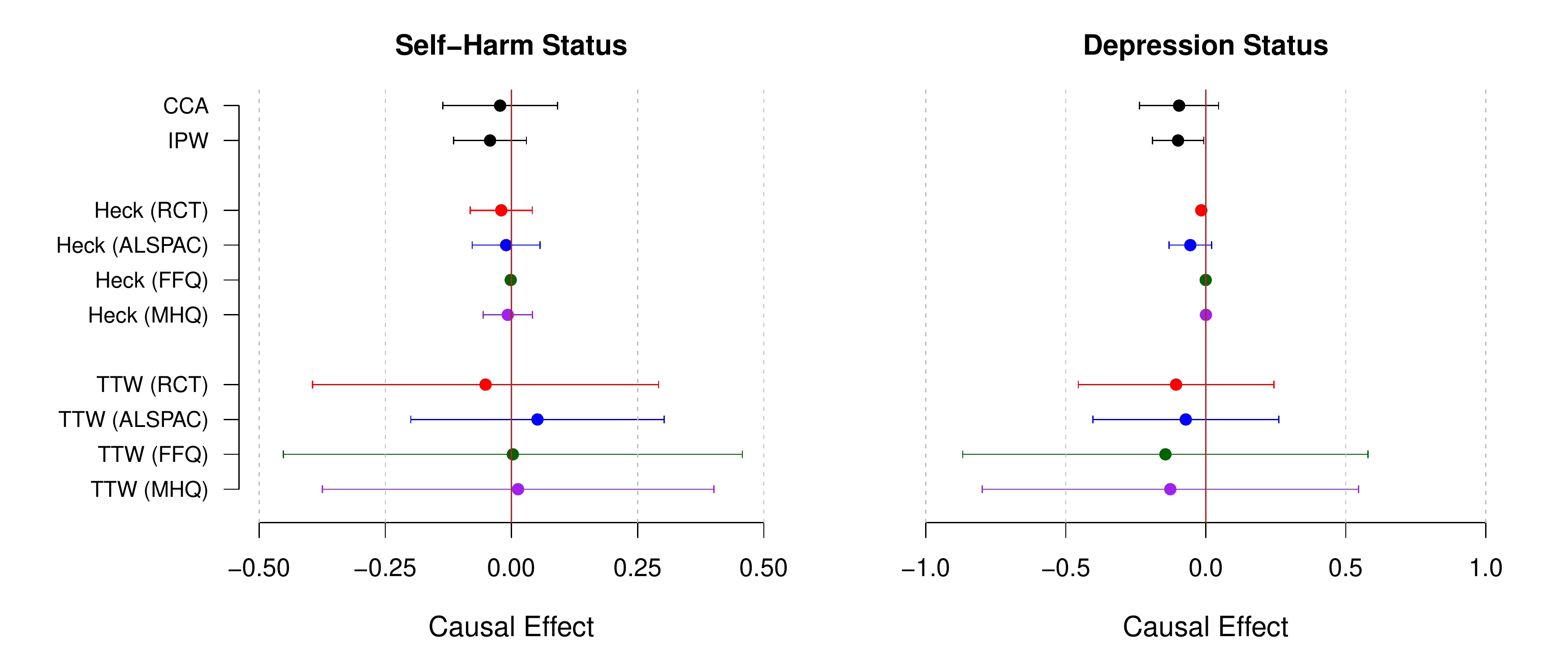}
\end{center}
\caption{MR estimates of the effect of BMI on self-harm (left) and depression status (right), obtained from ALSPAC data using various methods to adjust for missing data in BMI and the two outcomes. Colors indicate which instrument for selection is used each time (red: RCT for participation, blue: ALSPAC risk score, green: SNPs associated with FFQ completion in UK Biobank, purple: SNPs associated with MHQ completion in UK Biobank, black: method requires no instrument for selection).}
\label{alspac.supp.fig2}
\end{figure}

Data on self-harm and depression status were obtained from ALSPAC. Self-harm was self-reported by participants as part of the ``It's all about you" questionnaire (age 20+), while depression status was assessed during the TF4 clinic visit (age 18). Both were reported as binary variables. We estimated the effects of BMI on these traits using a Wald ratio estimate and the same polygenic risk score for BMI as in our analyses for smoking outcomes. We then implemented the four selection bias adjustment methods to assess whether our analyses may be biased due to missing data. Results are reported in Table~\ref{alspac.table3} and in graphical form in Figure~\ref{alspac.supp.fig2}.

\begin{center}
\begin{table*}[t]%
\caption{MR estimates of the effects of BMI on self-harm and depression status in ALSPAC, estimated using various methods and instruments for selection to adjust for missing data.\label{alspac.table3}}
\centering
\scriptsize
\begin{tabular}{cc@{\hskip 0.5in}ccc@{\hskip 0.5in}ccc}
  \hline
   \multirow{2}{*}{\textbf{Method}} & \multirow{2}{*}{\textbf{Instrument}} & \multicolumn{3}{c}{\textbf{Self-harm}} & \multicolumn{3}{c}{\textbf{Depression Status}} \\
   & & \textbf{Causal} & \textbf{SD} & \textbf{$95\%$ CI} & \textbf{Causal} & \textbf{SD} & \textbf{$95\%$ CI} \\ 
  \hline
  CCA 		& ----- 	& -0.022 & 0.058 & (-0.136 , 0.091) & -0.096 & 0.072 & (-0.237 ,  0.046) \\ 
  IPW 		& ----- 	& -0.042 & 0.037 & (-0.114 , 0.030) & -0.099 & 0.047 & (-0.191 , -0.008) \\ 
  Heckman 	& RCT 		& -0.020 & 0.031 & (-0.082 , 0.041) & -0.016 &  Inf  &      -----    \\ 
  Heckman 	& ALSPAC 	& -0.011 & 0.034 & (-0.078 , 0.057) & -0.056 & 0.039 & (-0.131 ,  0.020) \\ 
  Heckman 	& FFQ 		& -0.002 &  Inf  &      -----       & -0.001 &  Inf  &      -----    \\ 
  Heckman 	& MHQ 		& -0.007 & 0.025 & (-0.056 , 0.041) &  0.000 &  Inf  &      -----    \\ 
  TTW 		& RCT 		& -0.051 & 0.175 & (-0.394 , 0.291) & -0.107 & 0.178 & (-0.456 ,  0.243) \\ 
  TTW 		& ALSPAC 	&  0.052 & 0.128 & (-0.199 , 0.302) & -0.072 & 0.169 & (-0.404 ,  0.260) \\ 
  TTW 		& FFQ 		&  0.003 & 0.232 & (-0.452 , 0.458) & -0.144 & 0.369 & (-0.868 ,  0.579) \\ 
  TTW 		& MHQ 		&  0.013 & 0.198 & (-0.375 , 0.401) & -0.127 & 0.343 & (-0.799 ,  0.545) \\ 
  \hline
\end{tabular}
\end{table*}
\end{center}

Our analysis did not find evidence of a causal effect of BMI on self-harm and this was the case for all four selection bias adjustment methods. For depression status, IPW suggested a risk-decreasing effect, but that effect was quite weak and would not survive a correction for multiple testing. The other methods did not find evidence of a causal effect.

Once again, point estimates from the different methods were in agreement, and IVsel estimates using different instruments were also quite similar. The TTW method again produced estimates with high uncertainty compared to other methods.

Some implementations of Heckman's method suffered from collinearity, resulting in infinite standard error estimates. This was the case for the FFQ instrument in the self-harm MR analysis and for the FFQ, MHQ and RCT instruments in the MR analysis for depression status. For these analyses, we have only plotted the point estimates in Figure~\ref{alspac.supp.fig2}. In our experience, this performance of Heckman's method tends to occur in binary-outcome analyses with small sample sizes and weak instruments. The prevalence of each outcome may also influence the method's performance: only $11\%$ of genotyped ALSPAC participants who attended the clinic visit exhibited depression symptoms, compared to $60\%$ for smoking (for which Heckman's method produced reasonable standard error estimates).

Comparing our results to the relevant literature, \cite{Lim2020} has investigated the effects of genetically elevated BMI on the odds of individuals engaging in self-harm. The authors of that study were unable to identify a causal relationship, although a risk-increasing effect of extreme BMI on self-harm was observed, possibly suggesting that the relationship between BMI and self-harm is non-linear. Our findings are in agreement with the null effect reported in that paper, and the consistent results among different methods suggest that this null finding is unlikely to be due to selection bias. Concerning the relationship between BMI and depression status, the situation is less clear in the literature, with some authors suggesting no causal relationship \citep{Walter2015, Howe2021} and others reporting a risk-increasing effect \citep{Tyrell2018}. Our results in this applied analysis support a null effect, although as with smoking traits, we should acknowledge that our analyses were performed on fairly young individuals, that the sample size we used was much smaller than that used by other studies in the literature, and that our approach here cannot account for potential non-linearity in the BMI-depression effect.

\subsection*{Instrument Strength}

In the main part of our manuscript, we investigated the effect of instrument strength on the performance of IVsel methods, and illustrated the benefits of using strong instruments when implementing the methods in practice. Here, we report some measures of instrument strength for the four instruments used in our real-data application. In particular, we report the values of the $R^2$ and F statistics, which are commonly used to assess instrument strength. The $R^2$ statistic quantifies the proportion of variation in the trait studied (here, missingness) that is explained by each instrument. The F statistic is routinely used as a measure of instrument strength in instrumental variable analysis, with a common rule-of-thumb suggesting that an F statistic value lower than 10 indicates a weak instrument.

$R^2$ and F statistics for the association of the four instruments with missingness in BMI and the various outcomes were computed from linear regression models with robust standard errors. In addition, we computed the proportion of missing values for each trait, as well as the prevalence among observed values for the three binary traits. These figures are reported in Table~\ref{alspac.strength}. We use ``RCT" to denote the non-genetic instrument constructed from the RCT for questionnaire completion, ``ALSPAC" to denote the polygenic risk score for ALSPAC participation, ``FFQ" for the PRS constructed using variants associated with participation in the Food Frequency Questionnaire in UK Biobank and ``MHQ" for the PRS constructed from SNPs associated with completion of the Mental Health Questionnaire.

\begin{center}
\begin{table*}[t]%
\caption{Instrument strength diagnostics ($R^2$ and $F$ statistics) for the association of the various instruments for selection with BMI, smoking status, number of cigarettes smoked, self-harm and depression status in ALSPAC.\label{alspac.strength}}
\centering
\scriptsize
\begin{tabular}{cc@{\hskip 0.5in}ccccc}
  \hline
   \textbf{Instrument} & \textbf{Statistic} & \textbf{BMI} & \textbf{Smoking} & \textbf{Cigarettes} & \textbf{Self-harm} & \textbf{Depression} \\
  \hline
  \multirow{2}{*}{RCT} 		& $R^2$	&  0.01 \% 	&  0.09 \% 	&  0.07 \%	&  0.08 \% 	&  0.02 \% \\
  							& $F$ 	&   0.1 	&   4.8 	&   4.1 	&   4.3 	&   0.9 \\
  \multirow{2}{*}{ALSPAC} 	& $R^2$	&  4.53 \% 	&  3.50 \% 	&  3.50 \% 	&  3.59 \% 	&  4.05 \% \\
  							& $F$ 	& 409.4 	& 305.1 	& 308.2 	& 313.2 	& 363.7 \\
  \multirow{2}{*}{FFQ} 		& $R^2$	&  0.02 \% 	&  0.05 \% 	&  0.05 \% 	&  0.06 \% 	&  0.05 \% \\
  							& $F$ 	&   1.8 	&   4.1 	&   3.6 	&   3.6 	&   4.0 \\
  \multirow{2}{*}{MHQ} 		& $R^2$	&  0.08 \% 	&  0.26 \% 	&  0.27 \% 	&  0.28 \% 	&  0.10 \% \\
  							& $F$ 	&   6.4 	&  20.7 	&  21.5 	&  21.7 	&   8.0 \\
  \hline
  \multicolumn{2}{c}{\% Missing}	&  52.6 \% 	&  60.2 \%	&  60.6 \% 	&  60.0 \% 	&  56.8 \% \\
  \multicolumn{2}{c}{Prevalence}	&   --- 	&  60.8 \%	&   --- 	&  21.1 \%	&  11.3 \% \\
  \hline
\end{tabular}
\end{table*}
\end{center}

The ALSPAC-derived polygenic score was clearly the strongest instrument, explaining about $3.5\%$ of variation in selection for smoking traits and self-harm and an even higher proportion of variation in selection for depression and BMI. F statistics computed using this instrument were much higher than 10. Accordingly, in previous sections, IVsel estimates computed using the ALSPAC PRS as an instrument were subject to lower uncertainty than for the other three instruments.

The RCT, FFQ and MHQ instruments were all quite weak. For the RCT and FFQ instruments, the F statistics of association with selection in all five traits were below 10. The MHQ instrument exhibited acceptable association with selection for smoking traits and self-harm, but low association with selection for BMI and self-harm. The BMI results were particularly concerning, since BMI was used as a risk factor in all our MR analyses. Likewise, the values of the $R^2$ statistic were lower than $0.3\%$ for the MHQ instrument and all traits, and lower than $0.1\%$ for the RCT and FFQ instruments.

Missing data were common for all traits, with about $60\%$ of participants having missing data for smoking status, number of cigarettes and self-harm; the three traits were recorded as part of the same questionnaire, hence they share similar missingness patterns. BMI and depression status also shared similar missingness patterns, since they were recorded as part of the ``TF4" clinic visit. These two traits had missing values for approximately half the participants that passed our inclusion criteria (genotyped with recorded maternal traits).

\bibliography{ReFusion}%
\bibliographystyle{chicago}


